# Influence of surface relaxations on scanning probe microscopy images of the charge density wave material 2H-$NbSe_2$

Nikhil S. Sivakumar[1], Joost Aretz[1], Sebastian Scherb[1], Marion van Midden Mavrič[1,†], Nora Huijgen[1], Umut Kamber[1,‡], Daniel Wegner[1], Alexander A. Khajetoorians[1], Malte Rösner[1], Nadine Hauptmann[1*]

[1] Institute for Molecules and Materials, Radboud University, 6525 AJ Nijmegen, Netherlands

[†] Current address: Jožef Stefan Institute, Jamova 39, SI-1000 Ljubljana, Slovenia

[‡] Current address: Joseph Henry Laboratories and Department of Physics, Princeton University, Princeton, NJ, 08544, USA

[*] Correspondence to: n.hauptmann@science.ru.nl

**Scanning tunneling microscopy is the method of choice for characterizing charge density waves by imaging the variation in atomic-scale contrast of the surface. Due to the measurement principle of scanning tunneling microscopy, the electronic and lattice degrees of freedom are convoluted, making it difficult to disentangle a structural displacement from spatial variations in the electronic structure. In this work, we characterize the influence of the displacement of the surface-terminating Se atoms on the 3 x 3 charge density wave contrast in scanning probe microscopy images of 2H-$NbSe_2$. In scanning tunneling microscopy images, we observe the 3 x 3 charge density wave superstructure and atomic lattice at all probed tip-sample distances. In contrast, non-contact atomic force microscopy**

**images show both periodicities only at small tip-sample distances while, unexpectedly, a 3 x 3 superstructure is present at larger tip-sample distances. Using density functional theory calculations, we qualitatively reproduce the experimental findings and indicate that the 3 x 3 superstructure at different tip-sample distances in non-contact atomic force microscopy images is a result from different underlying interactions. In addition, we show that the displacement of the surface-terminating Se atoms has a negligible influence to the contrast in scanning tunneling microscopy images. Our combined experimental and theoretic work presents a method on how to discriminate the influence of the surface corrugation from the variation of the charge density to the charge density wave contrast in scanning probe microscopy images, which can provide insights into the influence of structural disorder to a charge density wave ground state.**

## I. INTRODUCTION

2H-$NbSe_2$ is a prototype material in which different phases with a collective ground state coexist. Next to a superconducting phase below 7.1 K [1], an almost commensurable 3 x 3 charge density wave (CDW) phase is formed below 33.5 K [2] to which we will refer as the CDW phase in the following. The coexistence of these phases has raised interest in understanding to what extent the underlying physical mechanisms interact [3-8]. The driving mechanism for the CDW in 2H-$NbSe_2$ has been discussed for more than two decades [3,9-12]. Following the Peierls model, a CDW in a one-dimensional material results from an electronic instability that modulates the charge density with a periodic lattice deformation mediated by electron-phonon coupling together with the opening of an energy gap [13,14]. Today, it is established that Fermi-surface nesting is not the main driver for the CDW formation in 2H-$NbSe_2$ [15-18]. Instead, momentum-dependent electron-phonon coupling is reported to be the driving mechanism for the CDW formation, based on various experimental and theoretical works [9,19-22]. In particular, the dependence of the electron–phonon coupling on the orbital character of the electronic states originating from Nb *d*-orbitals was shown to be an important factor for the CDW formation [23]. Recent works showed that the CDW remains present in the monolayer of 2H-$NbSe_2$, however, with drastically different transition temperatures [8,24]. This raises the question about the influence of the interface via the interaction of Se atoms with the supporting substrate. Furthermore, it has been recently suggested that the displacement of the surface Se atoms may play an important role in stabilizing the CDW [25], which necessitates methods to quantify the atomic-scale displacement of the Se atoms.

Scanning tunneling microscopy (STM) is an established method to characterize the CDW of 2H-$NbSe_2$ with atomic-scale resolution. STM probes the local density of states above the surface at the position of the tip within a given energy range. The CDW appears in constant-current STM images as a superposition of a nearly 3 x 3 superlattice together with the atomic lattice of the Se-terminated surface [26-28]. For simplicity, we refer to the nearly 3 x 3 CDW superlattice as the CDW in the following. The relative phase between the atomic lattice and the CDW has been shown to depend on the applied bias voltage, which was interpreted within a single-band CDW picture [26,29]. However, for a single CDW band around the Fermi level, an inversion of the contrast in STM images at small energies below and above the Fermi level is expected [30], which is not observed in 2H-$NbSe_2$ [26,31]. A previous work assigned this to strong vertical deformations of the Se atoms dominating the image contrast [32]. In a different work, the absence of a contrast in version in STM images around the Fermi energy was interpreted as a result from a multiband character of the CDW in 2H-$NbSe_2$ [31]. As STM is sensitive to the variations of the geometric height of the surface as well as to the local density of states at the tip position, it is challenging to separate a variation of the geometric height from atomic-scale variations of the electronic structure. To this end, complementary non-contact atomic force microscopy (nc-AFM) measurements provide a better characterization of the geometric structure [33,34] as nc-AFM is sensitive to the different interactions between the charge densities of the tip and the surface.

Here, we characterize the influence of the displacement of the surface-terminating Se atoms on the CDW contrast in STM and nc-AFM images of 2H-$NbSe_2$. We use distance-dependent current and frequency shift images to quantify the relative contributions of the atomic lattice and the CDW to the image contrast. Constant-height current images show both periodicities for the entire probed distance range. For constant-height frequency shift images, we observe, unexpectedly, the CDW contrast at large tip-surface distances, whereas the atomic lattice is also observed at smaller tip-surface distances. Using density functional theory (DFT) calculations within the Vienna Ab-Initio Simulation Package (VASP) based on a trilayer model for 2H-$NbSe_2$, we simulate current and frequency shift images and qualitatively reproduce the experimental observations. By comparing simulated images with and without a displacement of the surface Se atoms, we find that their displacement has a negligible influence on the CDW contrast in constant-height current images. In contrast, our calculations suggest that the CDW contrast in constant-height frequency shift images at different tip-surface distances result from different underlying interactions. At small distances, the CDW contrast is dominated by the Se-atom displacement. For larger tip-surface distances, we ascribe the dominant contribution to the CDW contrast to the interaction between the permanent dipole of the tip and the potential above the surface that is predominantly modulated by the underlying Nb atoms. In addition, a spatial variation of the work function may contribute to the CDW contrast at large tip-surface distances.

## II. METHODS

### A. Experiment

The experiments were conducted in a commercial ultra-high vacuum STM/AFM system [35] equipped with a Pan-style STM/AFM head. The setup operates at a base pressure of $\sim 2 \cdot 10^{-10}$ mbar and at a base temperature of $T \sim 6.2$ K. 2H-$NbSe_2$ crystals were bought from HQ Graphene [36] and cleaved in UHV at a temperature of $T \sim 160$ K. The 2H-$NbSe_2$ crystals were subsequently transferred *in situ* into the STM/AFM head. Combined STM/nc-AFM measurements were carried out utilizing a qPlus sensor [37] operated in non-contact frequency-modulation mode. The resonance frequency of the free oscillating prong was $f_0 \sim 27.7$ kHz and oscillation amplitudes $A_{\mathrm{osc}}$ = 30 pm to 60 pm (peak-to-peak value) were used. d$I$/d$V$ measurements were performed using a lock-in techniques. The bias voltage $V_{\mathrm{s}}$ is applied to the sample. We refer to Appendix A.1. for details about the tip material.

### B. Calculations

DFT calculations were performed within VASP 6.4.2 [38,39] using the Perdew-Burke-Ernzerhof (PBE) exchange correlation functional [40] and a projector augmented wave basis [41,42]. For self-consistent electronic structure calculations, $k$-meshes of (25 x 25 x 1) respectively (21 x 21 x 10) for n-layer and bulk were used. A plane wave energy cut-off of 366.5 eV was taken and Gaussian energy smearing with $\sigma = 0.03$ eV was applied. Structures of 1 to 4 layers of $NbSe_2$ in the 2H phase were relaxed until the norm of forces on each ion was smaller than $10^{-4} eV/$ Å using Methfessel-Paxton smearing with $\sigma = 0.2$ eV. Structural relaxations were done on smaller $k$-meshes but converged in the number of $k$-points. Typically, 5-7 $k$-points in the in-plane direction were sufficient. The CDW structures were obtained by structural relaxation of 3 x 3 in-plane supercells with respect to the 2H-$NbSe_2$ phase.

STM data is simulated using the Tersoff-Hamann model [43]. Describing the tip as a featureless point at position $\boldsymbol{r}$, and in the limit of a small applied bias voltage and low temperature, the tunneling current is approximated by

$$I_{\mathrm{sim}}(eV,\boldsymbol{r}) \propto \int_{\mathrm{E_f}}^{E_f+eV} dE \sum_{\lambda,k} T\left(k_{\lambda,y},k_{\lambda,x}\right)\left|\phi_{\lambda,\mathrm{k}}(\boldsymbol{r})\right|^2 \delta\left(\mathrm{E}-\mathrm{E_{k,\lambda}}\right) , \quad (1)$$

with $\left|\phi_{\lambda,k}(\boldsymbol{r})\right|^2$ the density at the tip position $\boldsymbol{r}$ of Kohn-Sham state in band $\lambda$ at momentum $k$ and tunneling matrix $T$ representing the tunneling probability into Kohn-Sham states in the sample with momentum $k$. The real-space pseudo-wavefunctions are reproduced using the PyVaspwfc package [44]. We model the larger sensitivity of the experimentally measured tunneling current to states with low in-plane momentum by letting

$$T\left(\,k_x,k_y\right) = \left(1/\left(\sigma_T\sqrt{2\pi}\right)\right) exp\left(\frac{-\left(k_x^2+k_y^2\right)}{\left(2\sigma_T^2\right)}\right), \quad (2)$$

with $\sigma_T = 0.2|b|$, where $b$ is the reciprocal in-plane lattice vector. We chose $\sigma_T$ such that the calculated DOS for bulk 2H-$NbSe_2$ is in good agreement with the experiment (see Sec. S9 in Supplemental Material [45]). Similar to the approach from Ref. [46], we simulate the differential conductance within the above-mentioned approximations via:

$$\frac{dI_{\mathrm{sim}}(eV,\boldsymbol{r})}{dV} \propto \sum_{\lambda,k} T\left(k_{\lambda,x},k_{\lambda,y}\right)\left|\phi_{\lambda,\mathrm{k}}(\boldsymbol{r})\right|^2 \delta\left(\mathrm{eV}-\mathrm{E_{k,\lambda}}\right), \quad (3)$$

neglecting the derivative of the tunneling matrix with the externally applied voltage, which is justified for small bias voltages. We simulate constant-height current ($I_{\mathrm{sim}}$) images at small positive bias voltages by including Kohn-Sham states between the Fermi energy $E_F$ and $E_F$ + 50 meV. In constant-height images of the differential conductance ($dI_{sim}/dV$), we include states around the simulated bias voltage $V_b$ within an energy window of $V_b \pm 25$ meV.

The DFT local potential including ionic and Hartree contributions was used to simulate the frequency-shift ($\Delta f_{\mathrm{sim}}$) following Ref. [47]. We model the tip as a point-like oscillator carrying a dipole and the force $F(z)$ as the interaction of this dipole with the Hartree and the ionic potential, calculated by:

$$F(\boldsymbol{r}) = -\nabla\left(\nabla v_{\mathrm{TS}}(\boldsymbol{r}) \cdot (\boldsymbol{p}_{\mathrm{per}} + \boldsymbol{p}_{\mathrm{ind}})\right). \quad (4)$$

Here $v_{\mathrm{TS}}\ (\boldsymbol{r})$ is the sum of Hartree and ionic potential resulting from the sample at the tip position $\boldsymbol{r}$ calculated with VASP. The first term is the contribution of the permanent dipole $\boldsymbol{p}_{\mathrm{per}}$ carried by the tip while the second term is the dipole which is induced by the electrostatic potential $\boldsymbol{p}_{\mathrm{ind}} = \tilde{\alpha}\nabla \cdot v_{\mathrm{TS}}(\boldsymbol{r})$ with $\tilde{\alpha}$ the tip-dependent polarizability. The permanent dipole is chosen to point perpendicular to the surface and the magnitude $\boldsymbol{p}_{\mathrm{per}}$ and $\tilde{\alpha}$ were set to unity. If the force field has a gradient along the path of oscillation this will shift the oscillation frequency of the oscillator by [48]:

$$\Delta f_{\mathrm{sim}}(x, y) = -\frac{f_0}{\pi A^2 k}\int_{-A}^{A} \frac{dF_z(x, y, z_0 - q)}{dz}\sqrt{A^2 - q^2}\,dq\,, \quad (5)$$

with $k$ being the stiffness of the tuning fork and $f_0$ its resonance frequency. An oscillation amplitude of $A = 50$ pm is considered as it was used in the experiment while the pre-factors in front of the integral are set to unity.

## III. RESULTS AND DISCUSSION

A representative constant-current STM topography of the 2H-$NbSe_2$ surface is shown in Fig. 1(a). We observe typical defects and vacancies as depicted in the large-scale STM topography in Fig. S1 of the Supplemental Material [45]. The atomic lattice and the CDW are clearly resolved in Fig. 1(a) as reported before [27,28], and can be identified as $q_{\mathrm{atom}}$ and $q_{\mathrm{CDW}}$ in the Fast-Fourier-Transformation (FFT) image, respectively (inset of Fig. 1(a)). We further observe a gradual spatial change in the apparent superstructure, going from a three-atom like feature (yellow circle), to a single-bright atomic feature (red dashed circle). This has been interpreted as the manifestation of the small deviation (~2%) of the CDW from a commensurate 3 x 3 superstructure [2,7]. The contrasts marked by the dashed and solid circles in Fig. 1(a) are typically associated with the chalcogen-centered (CC) and hollow-centered (HC) regions, respectively [4,7,49]. We note that the observed change in the apparent superstructure has recently been interpreted as the result of a competition of these two energetically-different structures [50] as well as domains of alternating triangular tiles of the two commensurate structures that are deformed by the presence of crystalline and topological defects [51].

In order to identify and characterize the accessible tip-surface distance range, we simultaneously measured the tunneling current $I_{\mathrm{t}}$ and frequency shift $\Delta f$ with respect to the tip displacement $z$ above a Se atom in the CC region (Fig. 1(b)). The tip displacement $z$ = 0 pm corresponds to a tunneling set point of ($V_{\mathrm{s}}$ = 100 mV and $I_{\mathrm{t}}$ = 10 pA) at which the constant-current feedback loop was opened. A smaller $z$ corresponds to a smaller tip-surface distance. For all used tips we find the same expected qualitative behavior of $\Delta f(z)$ and $I_{\mathrm{t}}(z)$ as depicted in Fig. 1(b). In particular, $\Delta f$(z) exhibits a minimum which indicates the onset of repulsive tip-surface interactions contributing to the total attractive force. Further details can be found in Appendix A.2. For distance-dependent images, we focus on tip-surface distances within $z$ = 0 pm and $z_{\mathrm{min}}$ which is the $z$ at the $\Delta f$(z) minimum.

To compare the atomic lattice and the CDW contrasts in STM and nc-AFM images, we acquired simultaneous images of $I_{\mathrm{t}}$ and $\Delta f$ in constant-height mode at different $z$ in a HC region (Figs. 1(c-f)). We refer to Appendix A.2 for further details on the measurement procedure and to Sec. S2 in the Supplemental Material [45] for the raw data at all probed tip-sample distances. In constant-height $I_{\mathrm{t}}$ images, we observe both, the atomic and the CDW contrast, at all probed tip-surface distances (Figs. 1(c, e)). This is also reflected by the finite intensity of $q_{\mathrm{atom}}$ and $q_{\mathrm{CDW}}$ in the FFT images (insets of Figs. 1(c, e)) which show a pattern similar to the FFT image in Fig. 1(a). We note that Figs. 1(c, d) are subject to a vertical drift (of $0.6 \pm 0.1$ pm/min). In contrast to the constant-height $I_{\mathrm{t}}$ image in Fig. 1(c), the simultaneously acquired $\Delta f$ image at $z$ = -50 pm (Fig. 1(d)) shows the CDW contrast without the atomic lattice, which is clearly depicted in the FFT image and the line profile (Fig. S3 of the Supplemental Material). We rule out any significant influence of crosstalk between $I_{\mathrm{t}}$ and $\Delta f$ to the CDW contrast in Fig. 1(d) as detailed in Sec. 3 of the Supplemental Material [45]. We find that the maxima in the CDW contrast in the constant-height $I_{\mathrm{t}}$ image correspond to the positions with the most negative $\Delta f$. This indicates a stronger attractive interaction at these positions. The observation of only the CDW contrast in $\Delta f$ images is unexpected under the assumption that the displacement of the surface Se atoms gives rise to the CDW contrast. The fact that we do not resolve the atomic lattice, indicates that short-range chemical interactions are not relevant to the image contrast at this tip-surface distance. Therefore, we suggest that the local displacement of the surface Se atoms cannot be the origin of the CDW contrast here.

Upon decreasing $z$ towards $z_{\mathrm{min}}$, the $\Delta f$ image shows the CDW contrast in addition to the atomic lattice (Fig. 1(f)), as shown in the FFT image in the inset and in the line profile in Fig. S3 of the Supplemental Material. A similar image contrast was obtained in previous contact AFM images of 1T-$TaSe_2$ and 1T-$TaS_2$ operating in the repulsive tip-sample interaction regime [52,53]. We therefore ascribe the observation of the atomic lattice to a stronger contribution of short-range chemical forces, *i.e.* forces resulting from Coulomb attraction and Pauli repulsion, to the total force given the fact that the images are acquired close to the minimum of $\Delta f$. For a CC region, we observe the same experimental findings for the simultaneously acquired $I_{\mathrm{t}}$ and $\Delta f$ images (see Sec. S4 in the Supplemental Material [45]). We note that the simultaneously acquired excitation signal of the tuning fork oscillation does not show any indication of dissipative processes for our probed tip-surface distances. Using inverse FFT images, we find the maxima in the CDW in the constant-height $I_{\mathrm{t}}$ image do not correspond to the positions with the most negative $\Delta f$ as observed for Figs. 1(c) and (d) (see Sec. S5 in the Supplemental Material [45]). We ascribe this to a stronger contribution of repulsive forces. Our observations indicate that the CDW contrasts in Figs. 1(d) and (f) result from different force contributions.

In order to elucidate the role of electrostatic forces on the CDW contrast in $\Delta f$ images, we performed Kelvin probe force microscopy (KPFM) [54,55]. We determined of the local contact potential difference (LCPD) between tip and sample by recording $\Delta f$ with respect to the sample voltage $V_{\mathrm{s}}$ at extreme positions of the CDW contrast in the CC and CH regions. Within the approximation of a plate capacitor for the tip-surface junction, $\Delta f(V_{\mathrm{s}})$ is expected to follow a parabolic shape, with the maximum voltage ($V_{\mathrm{LCPD}}$) corresponding to the LCPD [56,57]. A spatial variation of the electrostatic force component to the total force results into changes of the work function and is reflected by a spatial variation of $V_{\mathrm{LCPD}}$. The measurement procedure is detailed in Appendix A.3. Figure 2 shows $\Delta f(V_{\mathrm{s}})$ spectra averaged over 5 sweeps acquired at $z$ = -100 pm on the minimum (B) and the maximum (A1) positions of the CDW contrast in a CC area as depicted in the STM topography (inset). The $\Delta f(V_{\mathrm{s}})$ spectra show a parabolic shape, but we do not observe any variation of the $V_{\mathrm{LCPD}}$ within our error margin of 10 mV. $\Delta f(V_{\mathrm{s}})$ measurements in HC regions yield the same result. Furthermore, we do not find any variation of the $V_{\mathrm{LCPD}}$ within our error margin for $z$ > -100 pm. Given that this error margin translates to an uncertainty of the $\Delta f$ signal that is comparable to the CDW contrast variation in Fig. 1(d), we cannot exclude that a variation of the $V_{\mathrm{LCPD}}$ gives rise to the CDW contrast at large tip-sample distances. Thus, from our experiments, we cannot comment further on the origin of this unexpectedly observed CDW contrast. To this end, KPFM methods with a higher sensitivity could provide more insights in variations of $V_{\mathrm{LCPD}}$. This comprises to utilize functionalized tips that were shown to resolve anisotropic atomic charges [58], or heterodyne amplitude-modulation KFPM [59].

The relative intensities of the CDW contrast and the atomic lattice are different in the simultaneously acquired $I_{\mathrm{t}}$ and $\Delta f$ images at closer tip-surface distances (Figs. 1(e) and 1(f)). As it can be extracted from the FFT images, the CDW contrast is dominant in the $I_{\mathrm{t}}$ image whereas the atomic lattice dominates in the $\Delta f$ image. This is in line with the findings from previous works on 1T-$TaSe_2$ and 1T-$TaS_2$ [52,53,60]. The stronger CDW contrast in STM images was explained for 1T-$TaSe_2$ by a combination of the contribution of the charge density at the Fermi level and the vertical displacement of the Se atoms of ~30 pm [61,62]. Since the vertical displacement of Se atoms in 2H-$NbSe_2$ is only ~3 pm [20], the influence of the Se displacement to the STM image contrast may have a negligible contribution here.

In order to understand the origin of the CDW contrast observed in the different experimental images, we perform DFT calculations to compare qualitative trends in the distance-dependent image contrasts. We refer to Sec. II.b for the details on the calculation. In agreement with previous studies [4,49,63,64], we find two stable structural configurations which describe the HC and CC experimental regions, respectively (see Fig. S10 in the Supplemental Material [45].) These configurations alternate in ABAB fashion between layers. Figure 3 shows a comparison of the measured $dI_t/dV$ images for a CC region with $dI_{sim}/dV$ images for the trilayer model at a distance $d$ = 320 pm above the position of the surface-terminating Se atoms. For the images at $V_s$ = +40 mV (Figs. 3(a,d)) and $V_s$ = -50 mV ((Figs. 3(b, e)), we observe a good qualitative agreement between the measured and simulated images regarding the positions of the maximum contrast. For images at $V_s$ = -250 mV (Figs. 3(c, f)), we observe slightly larger deviations. We ascribe this deviation to a band close to the Γ-point with a dominant Se $p_z$ character, which has its band maximum around -250 meV for the trilayer model and around +300 meV in bulk, which is a result of enhanced interlayer hybridization and a subsequent strong $k_z$ dispersion in the bulk (see Sec. S8 in the Supplemental Material [45] for layer-dependent calculations of the band structure). For the HC region, we find a similar agreement between measured and simulated images (see Fig. S14 in the Supplemental Material [45]). Overall, we conclude that the trilayer model is sufficient to qualitatively describe the bulk properties of 2H-$NbSe_2$ in the energy window where we compare our calculations to the experimental data.

Next, we aim at understanding the measured distance-dependent contrast in constant-height $I_t$ and $\Delta f$ images. We simulate current ($I_{sim}$) and frequency shift ($\Delta f_{sim}$) images at different distances $d$ above the topmost Se atoms for the HC and CC configuration using the trilayer model. Figures 4(a, b) show the calculated $I_{sim}$ images for $d$ = 500 pm and $d$ = 300 pm, respectively, for the HC configuration. The atomic lattice and the CDW can be discerned at both $d$. This is similar for the simulated $\Delta f_{sim}$ images in Figs. 4(c, d). However, the relative contributions of the atomic and CDW contrast change with $d$. In Figs. 4(d), the atoms appear as clear protrusions which compares well with Fig. 1(f) and previous works [52,53]. At larger distances (Fig. 4(c)), the CDW contrast is stronger compared to the atomic lattice which is similar as in Fig. 1(c). However, different from Fig. 1(c), the atomic lattice is discernable in Fig. 4(c). We find similar trends for the CC regions (Fig. S15 in the Supplemental Material). We further note that the relative positions of the maxima of the CDW in $I_{sim}$ images and $\Delta f_{sim}$ images are different from the experimental findings and that the range for $d$ is larger than the range for $z$ in the experiment. We attribute these deviations to the approximations of our trilayer model and the highly simplified description of the tip which neglects any macroscopic contributions of van der Waals interactions as well as relaxations of the tip or surface. This is supported by our experimental findings using different tip materials and nanoscale tip shapes with different contributions of van der Waals interactions between tip and surface. We find that the absolute values strongly vary with different tips while the qualitative trends remain the same (Fig. S8 in the Supplemental Material). Therefore, we conclude that a qualitative comparison of the atomic and CDW contrasts of our theoretic model with the experimental findings is valid.

Next, we examine to what extent the CDW contrast in $\Delta f_{\text{sim}}$ images at large distances originates from the lateral or vertical displacements of the surface-terminating Se atoms. As extracted from the relaxed geometric structure in Sec. S8 in the Supplemental Material [45] and in agreement with previous works [20], the surface Se atoms are displaced by ~3 pm perpendicular to the surface forming a 3 x 3 superstructure. In addition, we find from our calculations that the Se atoms are also slightly displaced within the surface plane. To study the effects of the surface Se displacements, we further performed DFT calculations with these Se atoms fixed to their undistorted plane positions in the 2H phase using the trilayer model, along with adjusting the average Se-Nb distances to the undistorted one. The position of the plane of the topmost Se atoms was adjusted with respect to the Nb layer such that the Se-Nb distances fall within a range of 2.56 Å to 2.70 Å. This range is similar to the Se-Nb distance variation in the relaxed CDW structure and close to the distances for the undistorted $NbSe_2$ model (2.63 Å). We refer to this modified structural model as the fixed structure. The energy of the fixed structure (-534.43 eV) is higher than in the relaxed CDW structure (-534.56 eV), but lower than the structure without the CDW deformation (-533.35 eV). Figures 4(e, f) show the calculated $I_{\text{sim}}$ images for $d$ = 500 pm and $d$ = 300 pm, respectively, for the structure model with fixed Se atoms. We do not observe a significant modification to the image contrast as compared to the relaxed structure in Figs. 4(a, b). In contrast, for the calculated $\Delta f_{\text{sim}}$ images at $d$ = 300 pm (Fig. 4(h)), the CDW contrast appears to be strongly suppressed. Interestingly, the $\Delta f_{\text{sim}}$ image at $d$ = 500 pm (Fig. 4(g)) shows no obvious change with respect to the relaxed model, which indicates the images contrast to be insensitive to structural relaxations of the surface.

In order to compare the qualitative trends in the measured and calculated images for a larger range of tip-surface distances, we extract the FFT peak intensities ($A^{FFT}$) of $q_{atom}$ and $q_{CDW}$ from the different FFT images for the measured and calculated images. Figures 5(a, b) show $A^{FFT}$ $(z)$ extracted from the measured constant-height $I_t$ and $\Delta f$ images. Details of the analysis procedure can be found in Sec. S6 in the Supplemental Material [45]. For the constant-height $I_t$ images, we find a similar exponential trend of $A^{FFT}$ $(z)$ for $q_{atom}$ and $q_{CDW}$ (Fig. 5(a)), as indicated by the fitted dashed lines. We note that the absolute values and the exact slopes of $A^{FFT}(z)$ for the experimental data vary depending on the overall shape and material of the tip apex while the overall trends remain similar (see Fig. S8 in the Supplemental Material [45]). The experimental findings in Fig. 5(a) compare well to $A^{FFT}(d)$ for $q_{atom}$ and $q_{CDW}$ from the calculated $I_{sim}$ images, as shown in Figs. 5(c, e) for the relaxed and fixed structure, respectively. We note that the calculated values for d > 500 pm in Figs. 5(c, e) are within the noise levels due to the small energy range used for the current calculation ($E_F$ to $E_F$ + 50 meV). We disregard these data points in our discussion. We further note that the Γ-weighting in the $I_{sim}$ images has only a minor influence on the $A^{FFT}$ $(d)$ for $q_{CDW}$ and $q_{atom}$ within the energy range that we consider in the calculations.

For the measured constant-height $\Delta f$ images, we observe a smaller $A^{FFT}(z)$ for $q_{atom}$ than for $q_{CDW}$ for $z$ > -100 pm, while $q_{atom}$ becomes dominant at smaller $z$ (Fig. 5(b)). This behavior is also found in the calculated $\Delta f_{sim}$ images (Figs. 5(d, f)), where we observe that $A^{FFT}(d)$ for $q_{CDW}$ is larger than $q_{atom}$ for $d$ > 500 pm. We further find that $A^{FFT}(d)$ for $q_{atom}$ and $q_{CDW}$ in Figs. 5(d, f) does not follow a simple exponential trend. Our calculations show that for $d < 350$ pm, the tip-surface interaction is repulsive. At d > 350 pm, there are regions where the interaction of the dipole of our modeled tip with the sample is repulsive and regions where it is attractive, which gives rise to a kink in the $A^{\mathrm{FFT}}$ curve for $q_{\mathrm{atom}}$ (Figs. 4(d, f)) and to different slopes. This is different from the experimental $A^{\mathrm{FFT}}(z)$ for $q_{\mathrm{atom}}$ extracted from the measured $\Delta f$ image which shows a rather exponential behavior as indicated by the dashed line in Fig. 5(b). We assign this discrepancy to the missing van der Waals interaction between our modeled tip and the surface. Nevertheless, we conclude that our calculations qualitatively reproduced the distance-dependent contributions $A^{\mathrm{FFT}}$ to the image contrast of $q_{\mathrm{atom}}$ than for $q_{\mathrm{CDW}}$.

Next, we discuss the influence of the Se displacement on the CDW contrast in the different images. As seen from Figs. 5(c, e), the effect of the displacement of Se to $A^{\mathrm{FFT}}(d)$ for $q_{\mathrm{CDW}}$ and $q_{\mathrm{atom}}$ in the $I_{\mathrm{sim}}$ images is minimal for all tip-surface distances. We therefore conclude that the displacement of the surface-terminating Se atoms has no significant contribution to the CDW contrast in constant-height $I_{\mathrm{t}}$ images or STM topographies, as also shown in Fig. 4. Therefore, strong vertical deformations of the Se atoms are not the reason why STM images at small energy around the Fermi level do not show a contrast inversion as suggested in ref. [32]. This is different for the frequency shift images. As depicted in Fig. 5(f), the slope of $A^{\mathrm{FFT}}(d)$ for $q_{\mathrm{CDW}}$ is significantly modified for the fixed structure for $d \lesssim 300$ pm, while $A^{\mathrm{FFT}}(d)$ for $q_{\mathrm{atom}}$ shows only minor changes in relation to the relaxed structure in Fig. 5(d). We conclude that the displacement of the Se atoms is the dominant contribution to the CDW contrast in $\Delta f_{\mathrm{sim}}$ images at small $d$. Our findings are in line with previous results on $WS_2$ that utilized a CO-functionalized tip showing a strong sensitivity to surface deformation in $\Delta f$ images [65,66] and also with the contrast AFM images in contact mode of the CDW of 1T-$TaSe_2$ and 1T-$TaS_2$ [52,53,60]. Interestingly, the displacement of the surface-terminating Se atoms does not have a strong influence on the $A^{\mathrm{FFT}}(d)$ for $q_{\mathrm{CDW}}$ at $d > 500$ pm (compare Figs. 5(d, f)). We conclude that the CDW contrast in $\Delta f$ images at large distances does not result from the displaced Se atoms.

To understand the origin of the contrast at larger tip-sample distances, we first compare the contributions of the permanent ($\boldsymbol{p}_{\mathrm{per}}$) and induced ($\boldsymbol{p}_{\mathrm{ind}}$) dipole of the modeled tip to the total $\Delta f_{\mathrm{sim}}$ (see eq. (4) and (5)) at $d$ = 500 pm. We performed calculations for the relaxed and fixed structure where we neglected the contribution of $\boldsymbol{p}_{\mathrm{ind}}$ to $\Delta f_{\mathrm{sim}}$ ($\Delta f_{\mathrm{sim}}^{\mathrm{p}}$). As shown in Figs. 6(a, b), $\Delta f_{\mathrm{sim}}^{\mathrm{p}}$ images do not show a qualitative difference of the image contrast compared to the total $\Delta f_{\mathrm{sim}}$ images in Figs. 4(c, g). We conclude that $\boldsymbol{p}_{\mathrm{per}}$ is the dominant contribution to $\Delta f_{\mathrm{sim}}$ images at large tip-sample distances. We therefore neglect the contribution of the induced dipole and calculate (eq. (4)) the contribution $\delta(\Delta f_{\mathrm{sim}}^{\mathrm{p}})$ to the total $\Delta f_{\mathrm{sim}}$ which results from the difference in the potentials $\delta v_{\mathrm{TS}}(\boldsymbol{r}) = v_{\mathrm{TS}}^{\mathrm{CDW}}(\boldsymbol{r}) - v_{\mathrm{TS}}^{\mathrm{n}}(\boldsymbol{r})$ between the CDW phase ($v_{\mathrm{TS}}^{\mathrm{CDW}}(\boldsymbol{r})$) and 2H phase ($v_{\mathrm{TS}}^{\mathrm{n}}(\boldsymbol{r})$). The resulting $\delta(\Delta f_{\mathrm{sim}}^{\mathrm{p}})$ images in Figs. 6(c, d) show no qualitative difference between the fixed and relaxed structure indicating that $\delta(\Delta f_{\mathrm{sim}}^{\mathrm{p}})$ is not a result from the displacement of the surface-terminating Se atoms at large tip-surface distances. We therefore suggest that the displaced Nb atoms in the subsurface layer have a dominant contribution to $\Delta f_{\mathrm{sim}}$ at large tip-sample distances. We note, however, that a variation of the LCPD cannot be ruled out based on the LCPD measurements and that a spatial variation of the work function may contribute to the CDW contrast at large tip-surface distances.

In conclusion, we characterize the influence of the displacement of the surface-terminating Se atoms on the CDW contrast in STM and nc-AFM images of 2H-$NbSe_2$. Using combined STM/nc-AFM we quantify the distance-dependent contributions to the image contrast of the atomic lattice and CDW by means of the FFT intensities of constant-height current images and frequency shift images. For constant-height current images, we resolve the atomic lattice and CDW at all tip-sample distances. In contrast, for constant-height frequency shift images, we unexpectedly observe the CDW contrast at large tip-surface distances, while for smaller distances the atomic lattice is also resolved. Our DFT calculations qualitatively reproduce the experimental findings. From a comparison of the relaxed structure of our trilayer 2H-$NbSe_2$ model with a structural model where the displacement of the surface Se atoms is minimized, we conclude that the Se atom displacement has a negligible influence to the contrast in constant-height current images. We thereby eliminate that surface deformations are the reason for the lacking contrast inversion around the Fermi level in STM images. For constant-height frequency shift images, our calculations suggest that the CDW contrast results from different interactions depending on the tip-surface distance. At small distances, the displacement of the surface-terminating Se atoms has the strongest contribution. This is different for larger tip-surface distances where we show that the CDW contrast is not caused by the displacement of the outer Se atoms. We suggest that the dominant contribution to the CDW contrast results from the interaction between the permanent dipole in our modeled tip with the potential above the surface that is predominantly modulated by the underlying Nb atoms. However, from measurements of the local contact potential difference, we cannot rule out a contribution of a spatial variation of the work function to the CDW contrast. Our findings illustrate how combined STM/nc-AFM can be used to characterize the role of the surface corrugation to the CDW contrast. In the future, it will be interesting to characterize atomic-scale defects in CDW materials by means of combined STM and nc-AFM in order to distinguish, *e.g.*, the influence of structural disorder from electronic modifications on strongly-correlated electron ground states.

**Acknowledgements**

The experimental part of this project was supported by the European Research Council (ERC) under the European Union's Horizon 2020 research and innovation programme (Grant No. 947717). The computations were performed at the Dutch National Supercomputer Snellius under Project No. EINF-7490. J.A. and M.R. acknowledge support from the research program "Materials for the Quantum Age" (QuMat). The latter program (registration number 024.005.006) is part of the Gravitation program financed by the Dutch Ministry of Education, Culture and Science (OCW).

## APPENDIX A: EXPERIMENTAL DETAILS

### 1. Tip preparation procedure

We utilized qPlus sensors with W and Nb tip wires. Nb tips were prepared by AC etching (20% HCl solution, $V = 2$ V, $f = 1$ kHz) while W tips were treated with DC etching (1 mol NaOH, $V = 3.9$ V). After insertion into the STM/AFM head, the Nb tips were prepared by field emission and small voltage pulses. At first, we used a W(110) surface and typically obtained a frequency shift $\Delta f > -13$Hz at our benchmark constant-current feedback parameters ($V_s = 100$ mV, $I_t = 10$ pA). It was not possible to further reduce the overall $\Delta f$, and thereby the overall long-range force components, by field emission on the W(110) surface. Afterwards, we prepared both Nb and W tips on Au(111) surfaces by field emission, small voltage pulses and dipping into the surface until $\Delta f > -5$ Hz was achieved and until the d$I$/d$V$ ($V_s$) within a voltage range of $\pm$ 1V clearly shows the surface state of Au(111) and no additional resonances. We note that we did not observe any significant difference of the experimental findings when using different bulk materials of the tip.

## 2. Details on constant-height measurements

Prior to each set of constant-height images, the tunneling current $I_t$ and frequency shift $\Delta f$ with respect to the tip displacement $z$ were acquired above a Se atom in order to characterize the tip and the accessible tip-sample distance range, as shown in Fig. 1(b). We note that the tip displacement ($z_{min}$) and frequency shift ($\Delta f_{min}$) at the frequency shift minimum vary for different tips between $z_{min}$ = -198 pm and $\Delta z_{min}$ = -285 pm and $\Delta f_{min}$ = -16.5 Hz and $\Delta f_{min}$ = -34.9 Hz, respectively, which we assign to a different contribution of the van der Waals interaction to the total tip-surface interaction. The simultaneously measured $I_t(z)$ in Fig. 1(b) shows the expected exponential distance dependence (dashed line). For most of our tips, we observe a deviation from the exponential dependence on $z$ at smaller tip-surface distances. For the data in Fig. 1(b), we observe a deviation for $z > 200$ pm. The exact $z$ range depends on the tip shape as shown in Fig. S8 in the Supplemental Material [45], which we interpret as structural relaxation of the tip or the surface.

A constant-current STM topography was acquired prior to each set of simultaneously acquired $I_t$ and $\Delta f$ images to ascertain the lateral position. Subsequently, the tip was stabilized at a fixed position on top of a Se atom in the CC region (stabilization parameters $V_s$ = 100 mV, $I_t$ = 10 pA) and the constant-current feedback-loop was opened. Constant-height images of $I_t$ and $\Delta f$ were acquired simultaneously at the respective $z$. Subsequently, the current feedback-loop was closed and a constant-current STM topography was acquired to account for lateral drift. Afterwards, the procedure was repeated for different $z$ within the range of $z$ = 0 pm and $z = z_{min}$.

### 3. LCPD spectra

First, a constant-current STM image was obtained to identify the maximum and minimum position. Then, the tip was stabilized on top of the maximum position and the current feedback-loop was opened at $V_s$ = 100 mV $I_t$ = 10 pA. LCPD measurements, *i.e.*, $\Delta f$ versus $V_s$ sweeps, were performed at constant height in CC regions on the maximum position (A1), followed by the minimum position (B) and then repeated at the maximum position (A2) to exclude any tip modifications between the measurements or any lateral or vertical drift. Afterwards, the current feedback-loop was closed again and an STM topography was acquired to check for any lateral drift. These measurements were repeated multiple times for $z$ = 0 pm, $z$ = -50 pm and $z$ = -100 pm. We extract the maximum positions by a parabolic fit to be $V_{LCPD}^{A1}$ = 0.837 ± 0.007 V and $V_{LCPD}^{B}$ = 0.840 ± 0.004 V. The control measurement at the maximum position A results in $V_{LCPD}^{A2}$ = 0.840 ± 0.006 V. We obtain similar values for HC regions, but the error margin tends to be slightly higher due to a larger uncertainty in the spatial positioning of the tip. Overall, we obtain an averaged uncertainty margin of about 10 mV. We conclude that we do not observe a variation of the LCPD larger than 10 mV in neither the CC nor HC region for $z$ > -100 pm.

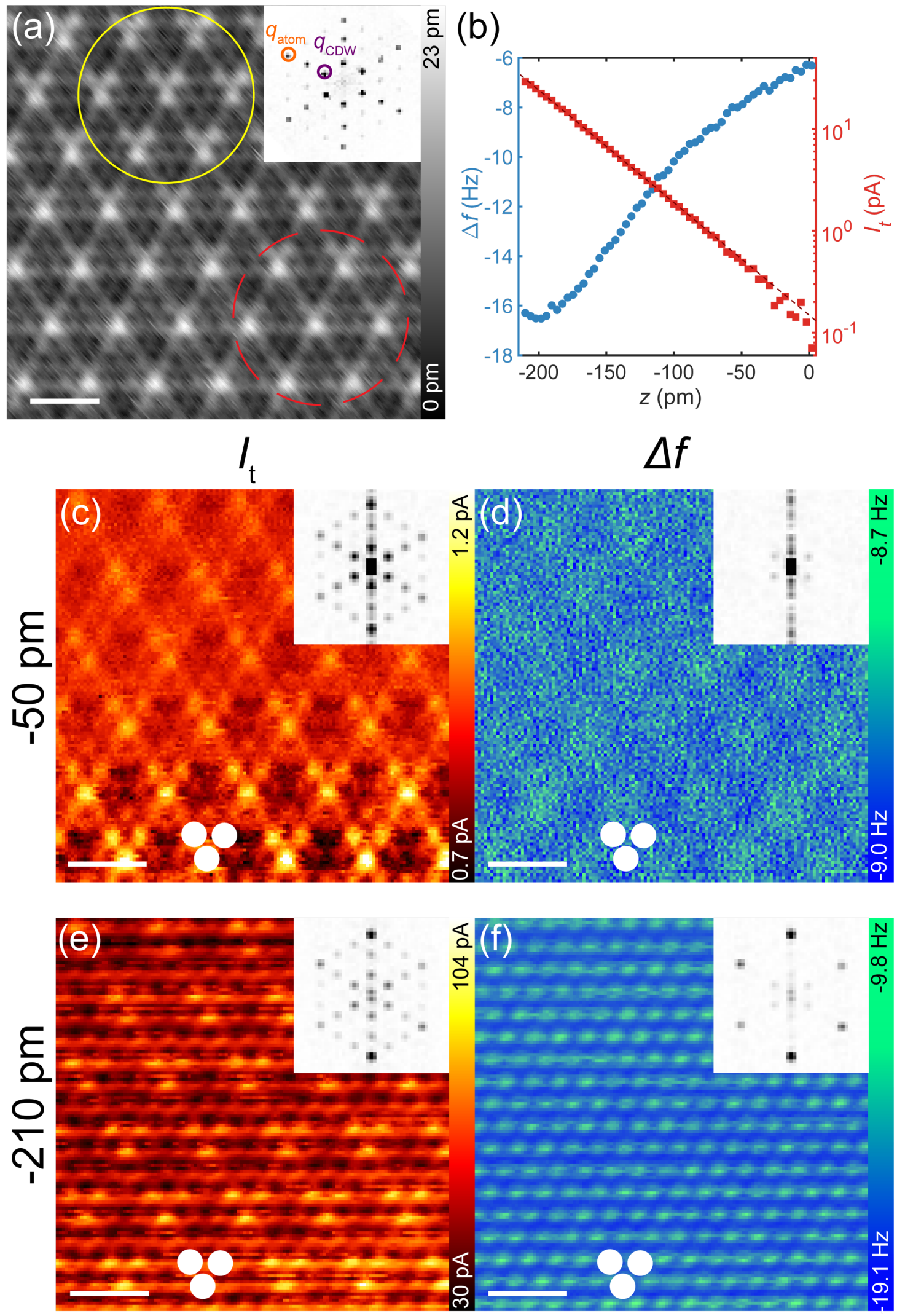

Figure 1: (a) Constant-current STM topography of 2H-$NbSe_2$ ($V_s$ = 100 mV, $I_t$=10 pA, line-flattened). (b) Frequency shift $\Delta f$ and current $I_t$ versus tip displacement $z$. More negative $z$ correspond to smaller tip-surface distances. Constant-height (c) $I_t$ and (d) $\Delta f$ images at $z$ = -50 pm acquired in a HC region ($V_s$ = 3 mV, $A_{osc}$ = 50 pm, both images line-flattened). (e) $I_t$ and (f) $\Delta f$ images at $z$ = -210 pm ($V_s$ = 3 mV, $A_{osc}$ = 50 pm, both images plane-corrected). Insets in (a, c-f) show the respective FFT images. The tip has been stabilized at $V_s$ = 100 mV, $I_t$ = 10 pA for the constant-height images (c-f). Scale bars are 1 nm in all images. The white circles in (c-f) mark the positions of the brightest appearing Se atoms in the current images.

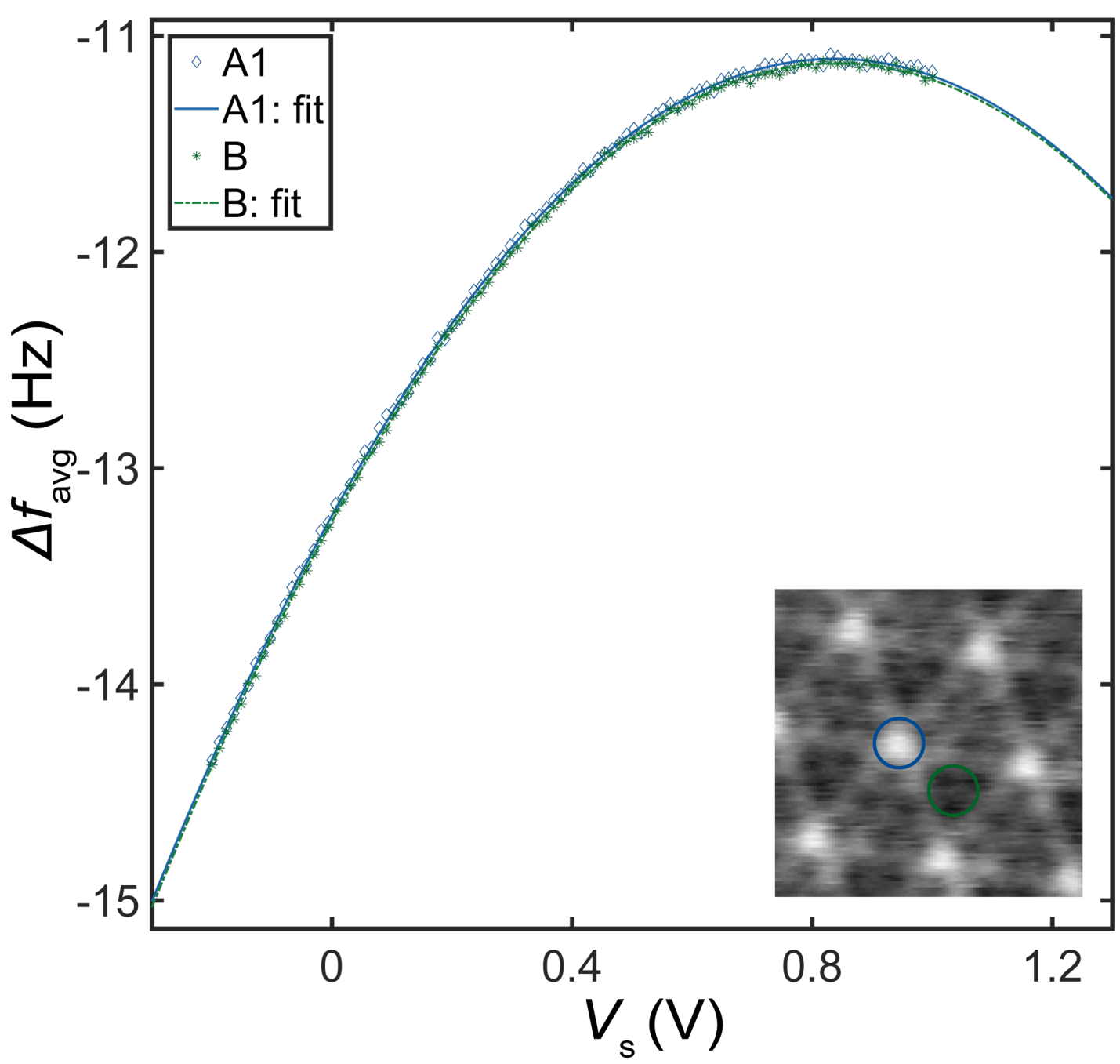

Figure 2: $\Delta f\,(V_s)$ spectra averaged over 5 sweeps obtained a minimum (B, green) and maximum (A1, blue) of the CDW. Inset: STM topography of a CC region with the circles indicating the measurement positions at minimum and maximum contrast of the CDW ($V_s$ = 100 mV, $I_t$ = 100 pA, $A_{osc}$ = 60 pm, $z$ = -100 pm).

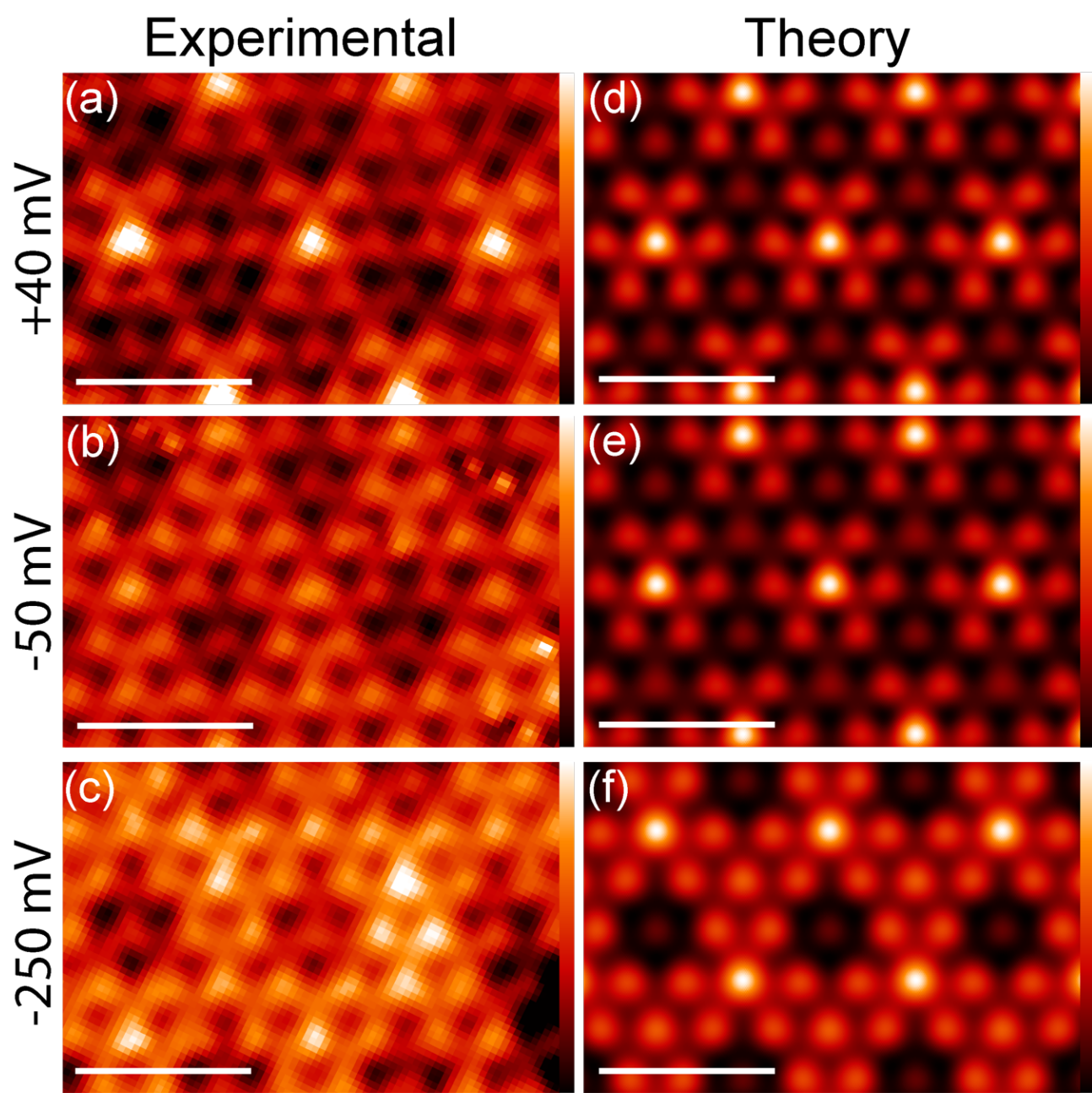

Figure 3: (a-c) Measured $\mathrm{d}I_t/\mathrm{d}V$ and (d-f) simulated $\mathrm{d}I_{sim}/\mathrm{d}V$ images at (a,d) $V_s = +40$ mV, (b,e) $V_s = -50$ mV and (c,f) $V_s = -250$ mV. The simulated images are calculated at $d = 320$ pm above a CC region. Tip stabilization: $V_s = 50$ mV, $I_t = 100$ pA , $V_{mod} = 5$ $\mathrm{mV_{rms}}$ above a CC region. (a-c) are line-flattened and Gaussian-filtered with 5 points, and the color scale is adjusted for maximum contrast. Scale bars are 1 nm.

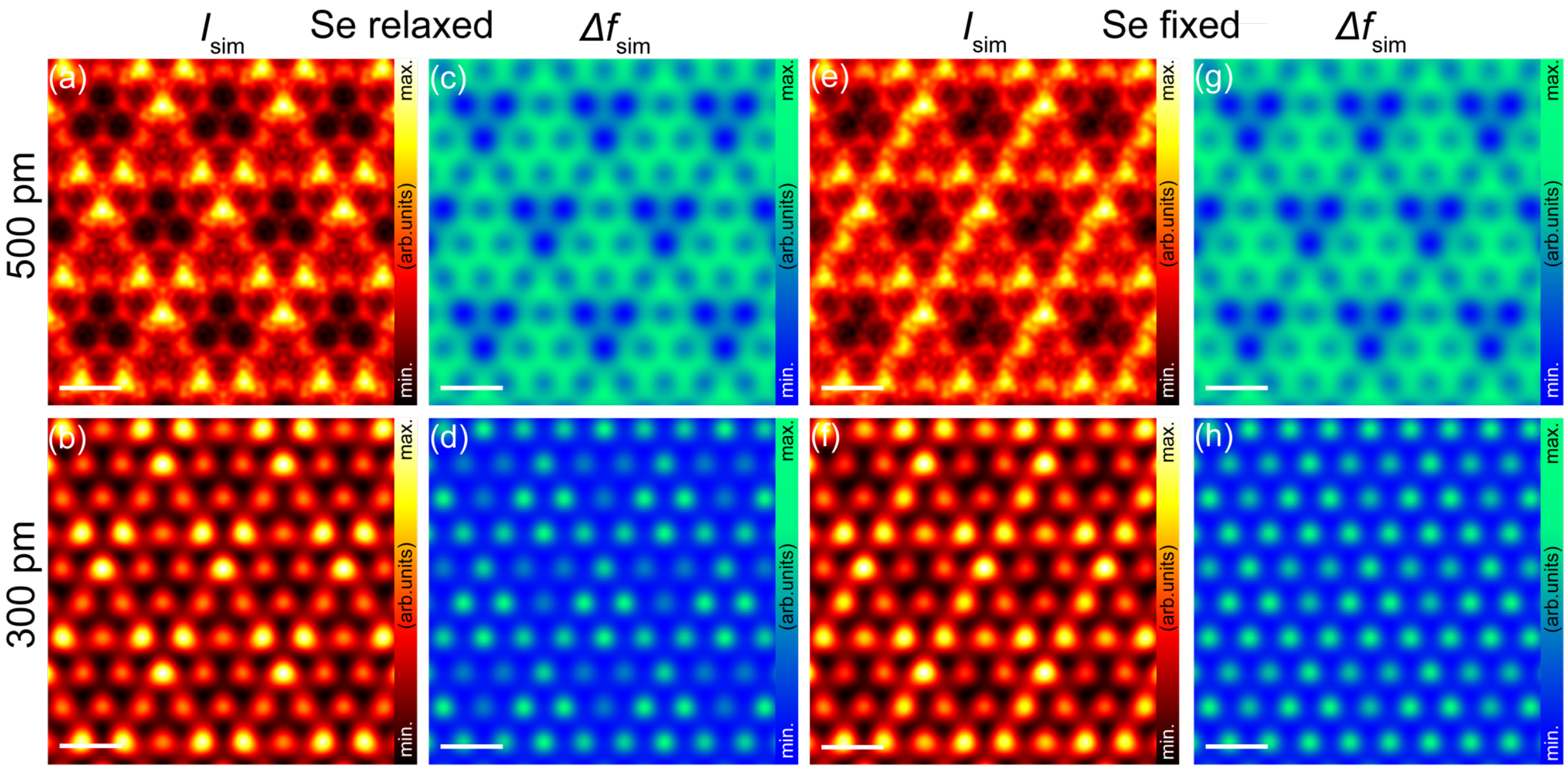

Figure 4: (a, b) Simulated current $I_{sim}$ and (c, d) simulated frequency shift $\Delta f_{sim}$ images for a HC region for different distances $d$ for the relaxed structure. (e-h) Same as in (a-d) but for the fixed structure. The current images were calculated using energy states between $E_F$ and $E_F + 50$ meV. We note that the minimum contrast value in (g) is negative. Scale bar is 0.5 nm in all images.

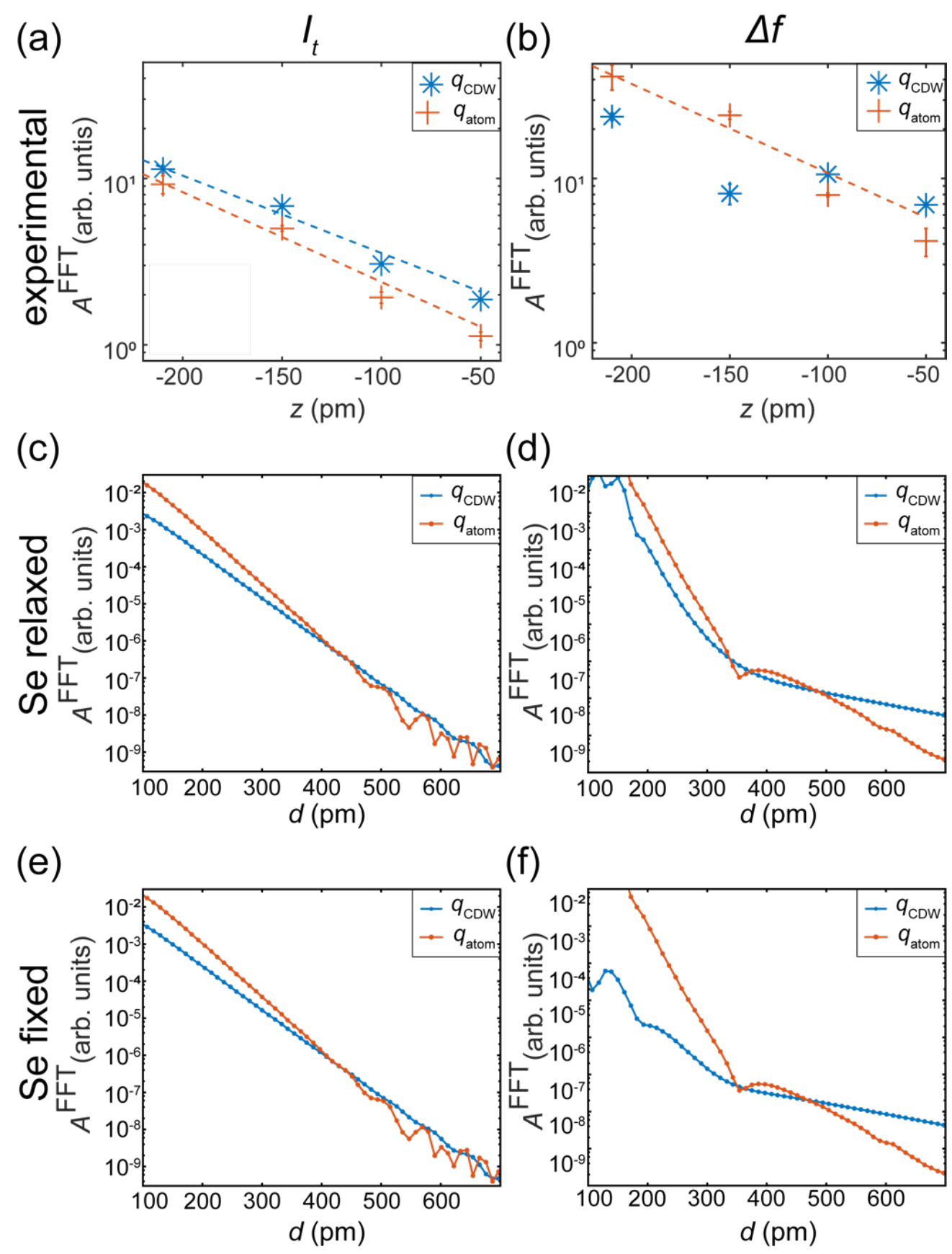

Figure 5: (a, b) Semi-log plot of FFT peak intensity ($A^{\mathrm{FFT}}$) of the atomic lattice ($q_{\mathrm{atom}}$) and the CDW contrast ($q_{\mathrm{CDW}}$) versus the tip displacement $z$, extracted from constant-height $I_{\mathrm{t}}$ (a) and $\Delta f$ images (b). (c, d) $A^{\mathrm{FFT}}$ of $q_{\mathrm{atom}}$ and $q_{\mathrm{CDW}}$ vs. $d$ extracted from $I_{\mathrm{sim}}$ (c) and $\Delta f_{\mathrm{sim}}$ (d) images for the relaxed structure. (e, f) Same as (c, d) but for the structure with fixed Se atoms. All images were measured or calculated for a HC region. We note that the calculated values for d > 500 pm in panels (c) and (e) are within the noise levels of our calculation and are disregarded in our discussion.

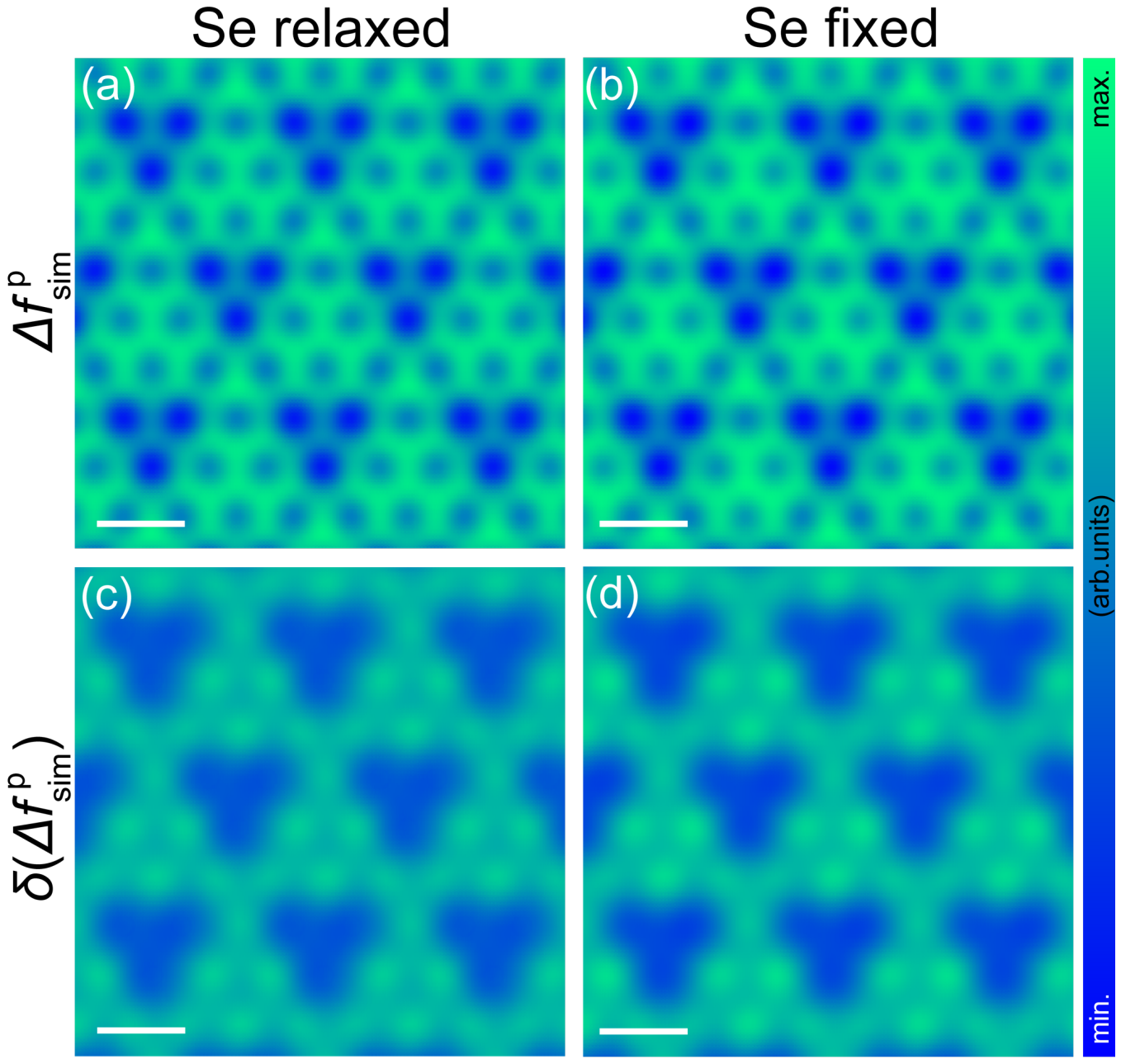

Figure 6: (a, b) Simulated images of the frequency shift $\Delta f^{\mathrm{p}}_{\mathrm{sim}}$ resulting from only the interaction of the permanent dipole of the tip with the surface potential for the relaxed and fixed structure at $d$ = 500 pm. (c, d) Difference of the frequency shift $\delta(\Delta f^{\mathrm{p}}_{\mathrm{sim}})$ resulting from the difference $\delta v_{\mathrm{TS}}(\boldsymbol{r})$ in between the potentials of the structures in the CDW and 2H phase for the relaxed and fixed structure at $d$ = 500 pm. All images were calculated for a HC configuration. Scale bar is 0.5 nm in all images.

# Supplemental Material: Influence of surface relaxations on atomic-resolution imaging of a charge density wave material

Nikhil Seeja Sivakumar[1], Joost Aretz[1], Sebastian Scherb[1], Marion van Midden Mavric[1,†], Nora Huijgen[1], Umut Kamber[1,‡], Daniel Wegner[1], Alexander A. Khajetoorians[1], Malte Rösner[1], Nadine Hauptmann[1*]

[1] Institute for Molecules and Materials, Radboud University, 6525 AJ Nijmegen, Netherlands

[†] Current address: Jožef Stefan Institute, Jamova 39, SI-1000 Ljubljana, Slovenia

[‡] Current address: Joseph Henry Laboratories and Department of Physics, Princeton University, Princeton, NJ, 08544, USA

[*] Correspondence to: n.hauptmann@science.ru.nl

## Section S1: Surface characterization

Figure S1 shows a large-scale STM topography image of a representative 2H-$NbSe_2$ surface. Next to the hexagonal Se terminated atomic lattice and CDW superlattice as detailed in the main manuscript, typical defects and vacancies are observed which appear as dark spots or bright protrusions, in agreement with previous observations [1-4].

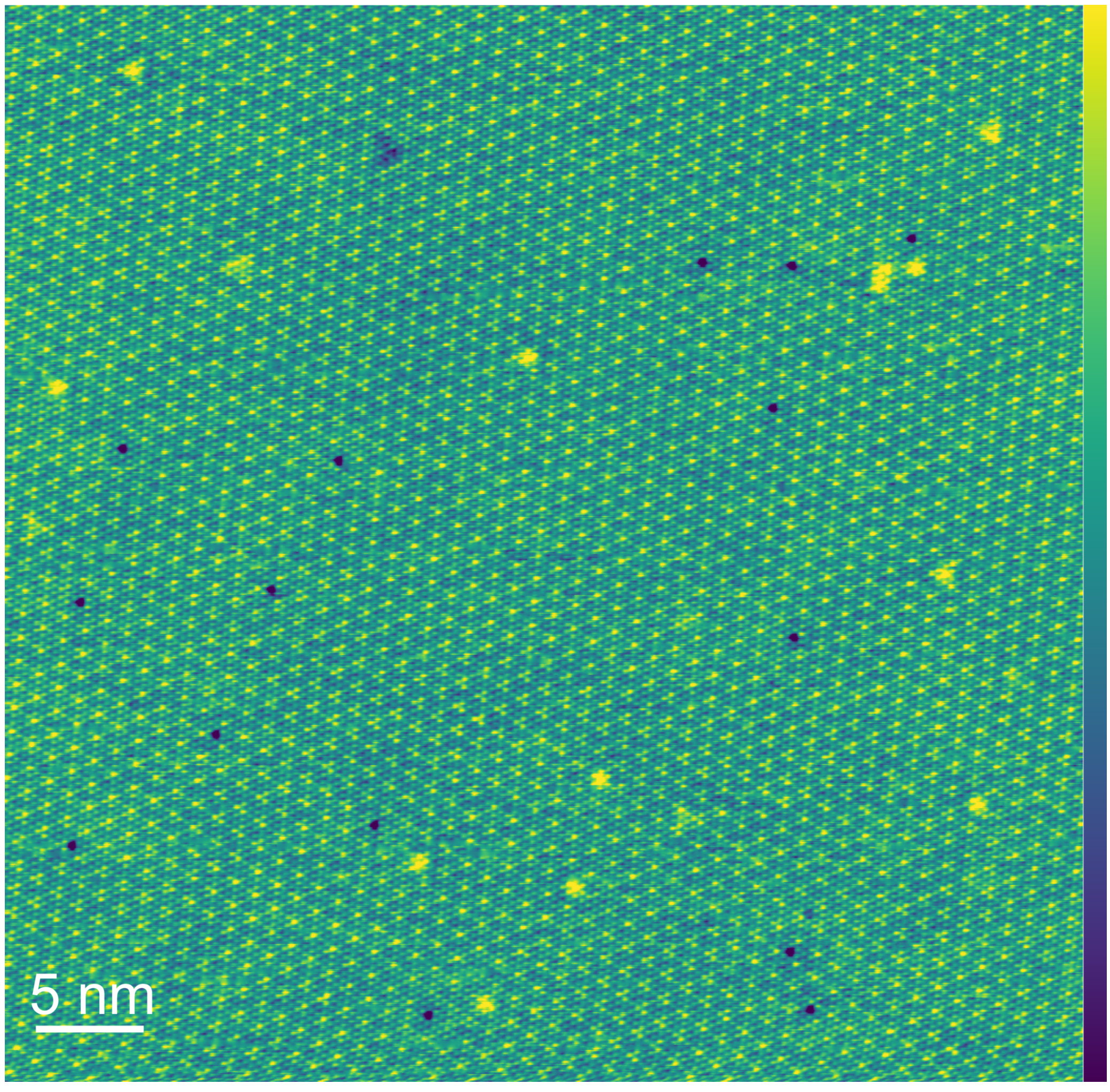

Figure S1: Typical large-scale STM topography of 2H-$NbSe_2$ ($V_s$ = 50 mV, $I_t$ = 100 pA, line-flattened).

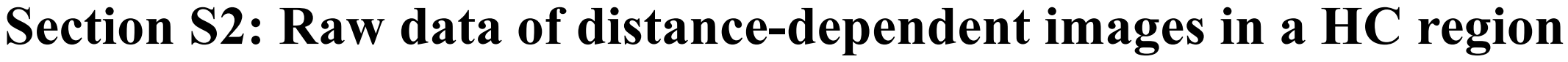

## Section S2: Raw data of distance-dependent images in a HC region

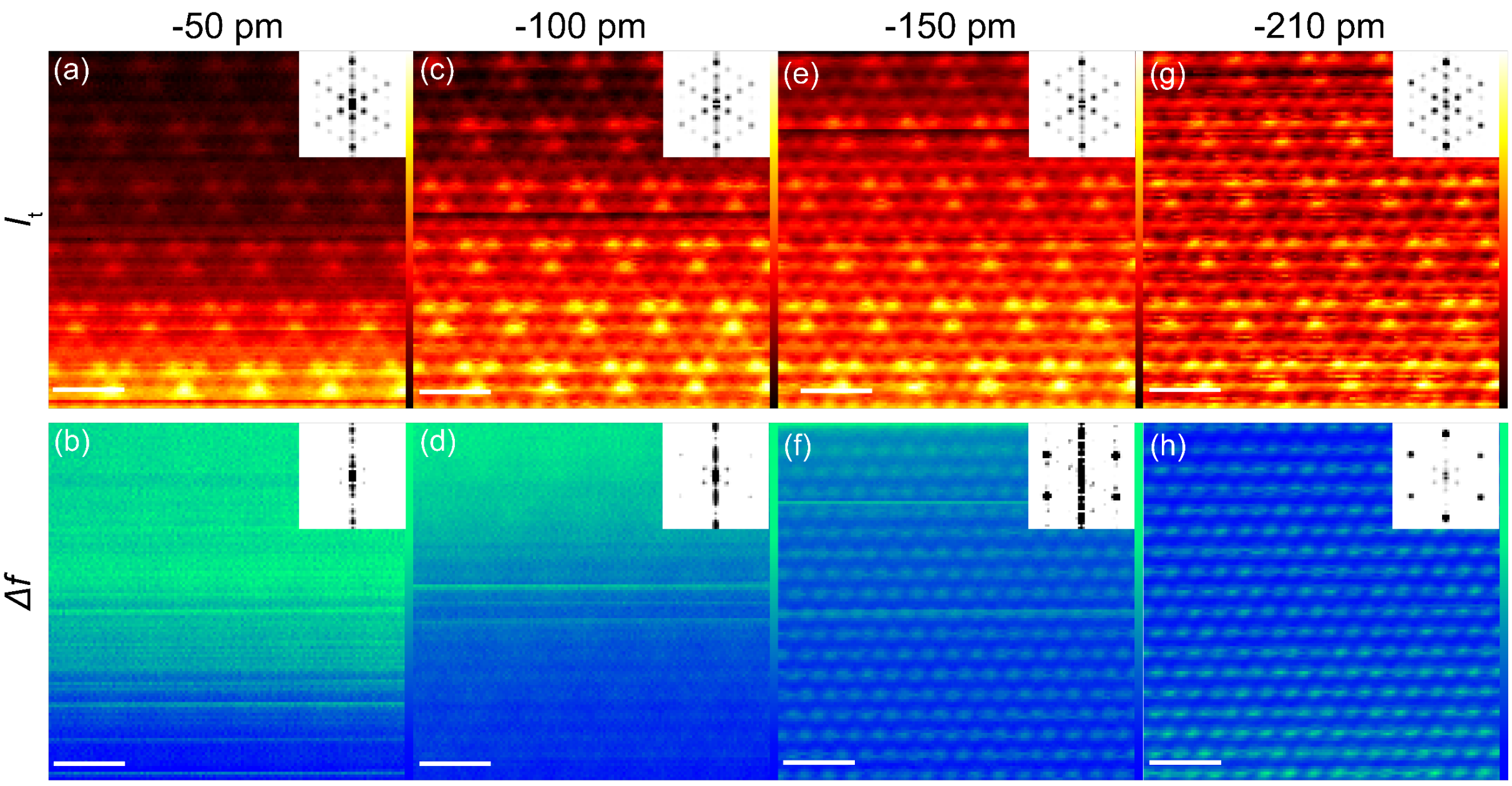

Figure S2: Raw data of simultaneously acquired constant-height current $I_t$ and frequency shift $\Delta f$ images at indicated tip displacements $z$. Insets: FFT images the respective $I_t$ and $\Delta f$ images. For all images: $V_s$ = 3 mV, $A_{osc}$ = 50 pm, tip stabilization: $V_s$ = 100 mV, $I_t$ = 10 pA. Scale bars are 1 nm. Color scales are (a) 0.4 to 3.2 pA, (b) -1.6 to -0.1 Hz, (c) 1.9 to 6.9 pA, (d) -12.7 to -10.2 Hz, (e) 4.4 to 31 pA, (f) -17.2 to -12.3 Hz, (g) 24 to 107 pA, (h) -17.7 to -7.7 Hz.

Figure S2 shows the full set of raw data for the constant-height $I_t$ and $\Delta f$ images at different tip displacements $z$ from Figs. 1(c-f) of the main manuscript. In all the $I_t$ images (Figs. S2(a), (c), (e), (g)) the hexagonal atomic lattice and the CDW corrugation are visible with similar intensity. The $\Delta f$ images only show atomic contrast at smaller tip-surface distances (corresponding to a more negative $z$) while the atomic lattice is absent in Fig. S2(b) and solely the CDW is discernible with a faint contrast. In Fig. S2(d), the atomic lattice starts to become observable and dominates the contrast in Figs. S2(f) and S2(h). This is more clearly illustrated by the inset FFT images, which show both periodicities for $I_t$ images while the atomic lattice becomes more prominent in $\Delta f$ images for more negative $z$. We note that in the inset of Fig. S2(f) only two high-symmetry axes of the CDW periodicity are visible. This is due to two factors: (i) the weak intensity of the CDW and (ii) the orientation of the fast scan direction with respect to the FFT peaks on the high-symmetry axis. In further obtained data sets, the image acquisition was rotated such that all three symmetry axes could be clearly observed.

## Section S3: Discussion about crosstalk between current and frequency shift

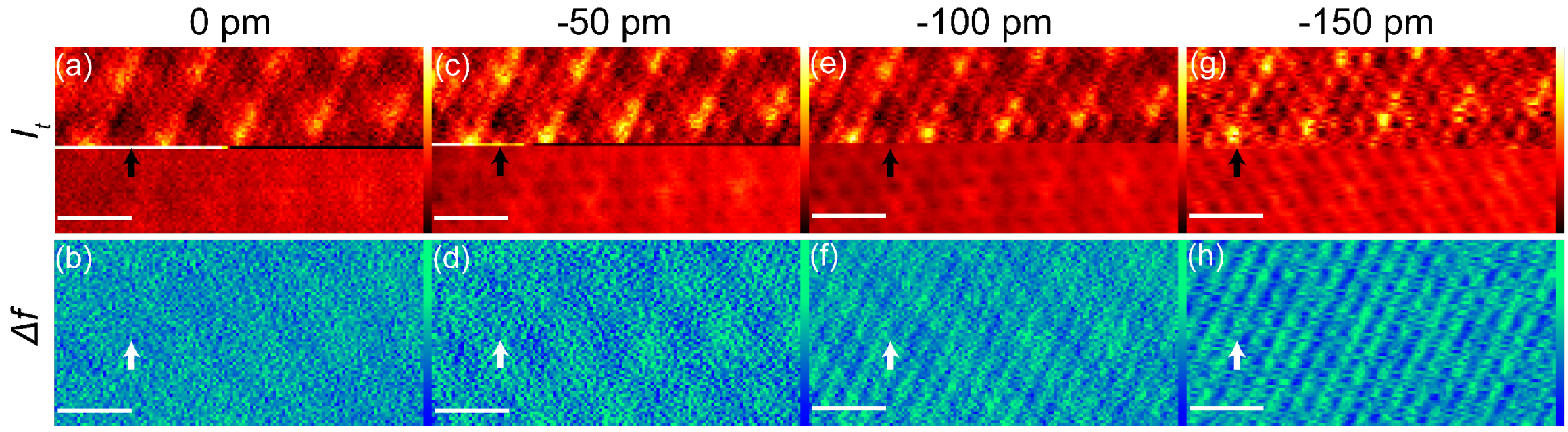

Figure S3: Constant-height current $I_t$ and frequency shift $\Delta f$ images at indicated tip displacements $z$ ($V_s$ = 3 mV, $A_{osc}$ = 30 pm). $V_s$ was set to 0 V during the scan as indicated by the arrows. Tip stabilization: $V_s$ = 100 mV, $I_t$ = 10 pA. Scale bars 1 nm. Color scales are (a) 0.6 to 1.1 pA, (b) -12.7 to -12.2 Hz, (c) 2.2 to 3.7 pA, (d) -16.8 to -16.4 Hz, (e) 5.4 to 11 pA, (f) -21.7 to -20.8 Hz, (g) 37 to 90 pA, (h) -27.6 to -23.0 Hz.

The qPlus sensor used in this work has previously been shown to be susceptible to crosstalk between $I_t$ and $\Delta f$ at larger currents [6-8]. To ascertain that the observed contrast in the $\Delta f$ images is not driven or influenced by capacitive coupling to the increasing current, we performed constant-height $I_t$ and $\Delta f$ imaging where the $V_s$ was set to 0 V halfway through the image. Otherwise, the measurement procedure was the same as described in Appendix A.2 in the main manuscript. As it can be seen in Figs. S3(a), (c), (e), and (g), the constant-height $I_t$ images show a clear expected change of contrast when $V_s$ is modified while the constant-height $\Delta f$ maps do not depict a significant influence. In particular, the faint CDW contrast is not altered. If the contrast in constant-height $\Delta f$ maps was driven or largely influenced by crosstalk from $I_t$, the contrast in the $\Delta f$ maps would have decreased simultaneously with the change of $V_s$. We therefore exclude a significant influence of crosstalk from $I_t$ to the patterns that we observe in the constant-height $\Delta f$ images.

## Section S4: Raw data of distance-dependent images in a CC region

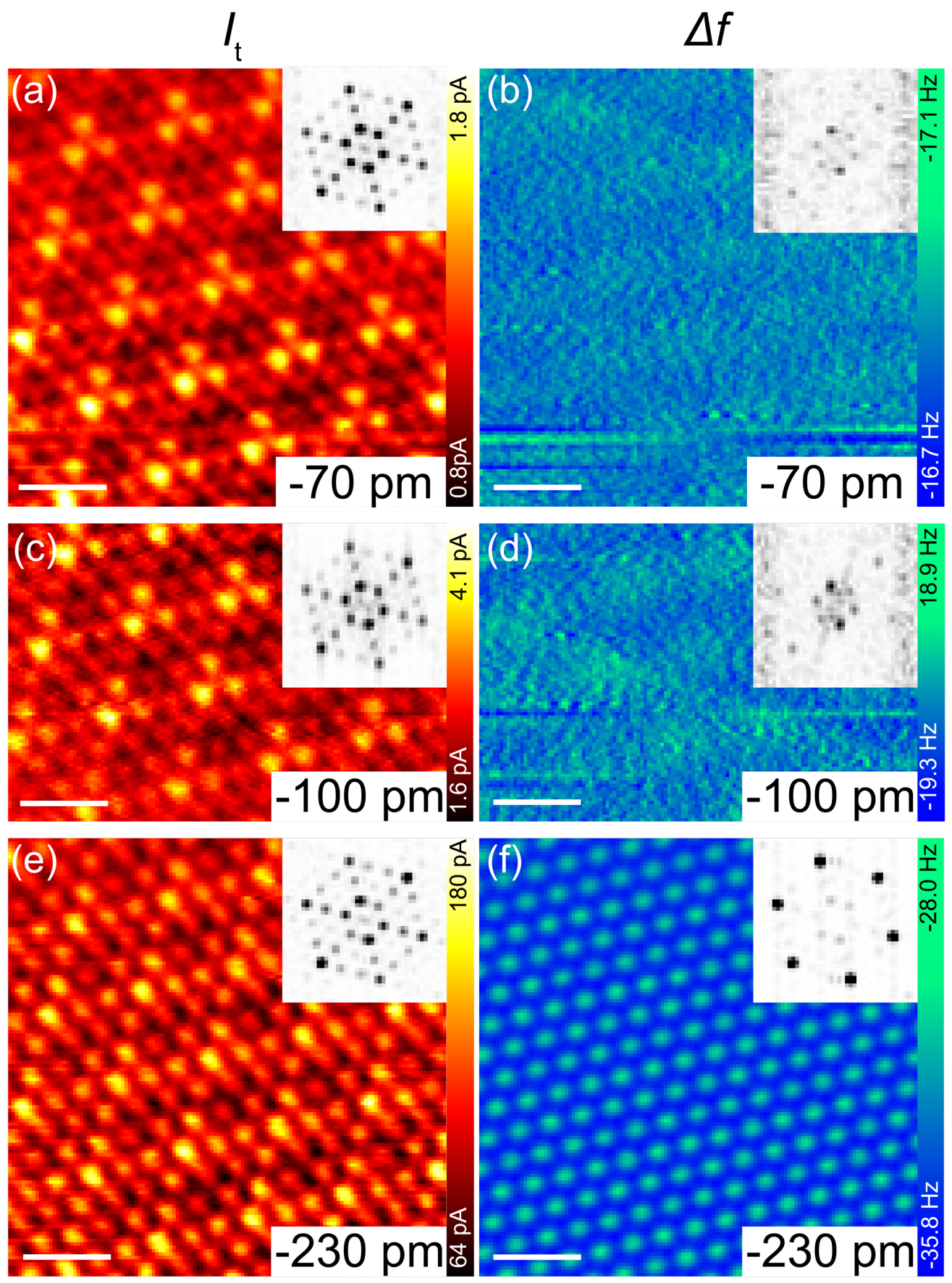

Figure S4: Constant-height (a) $I_t$ and (b) $\Delta f$ images at $z$ = -100 pm ($V_s$= 3 mV, $A_{osc}$ = 60 pm, $I_t$ and $\Delta f$ images row corrected and $\Delta f$ image gaussian filtered with 3 points). (c) $I_t$ and (d) $\Delta f$ images at $z$ = -230 pm ($V_s$ = 3 mV, $A_{osc}$ = 60 pm, $I_t$ and $\Delta f$ images row corrected). Insets show the respective FFT images. The tip has been stabilized at $V_s$ = 100 mV, $I_t$ = 10 pA for the constant-height images. Scale bars are 1 nm in all images.

Figure S4 shows constant-height $I_t$ and $\Delta f$ images at different z for a region closer to the CC structure. We note that we barely found regions of only CC or HC due to the incommensurably of the CDW. We observe the same trend for CC regions as observed for the HC region. $I_t$ images show contrast of both the atomic and CDW for all $z$ while the $\Delta f$ images at large tip-sample distances ($z$ = -70 pm) dominantly show a faint CDW. At smaller tip-sample distances, both lattices are observed in $\Delta f$ images.

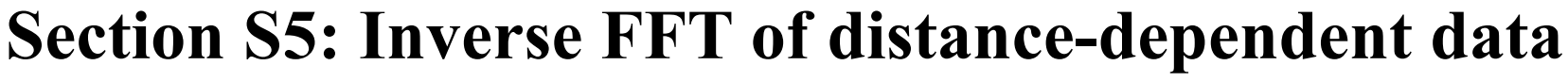

## Section S5: Inverse FFT of distance-dependent data

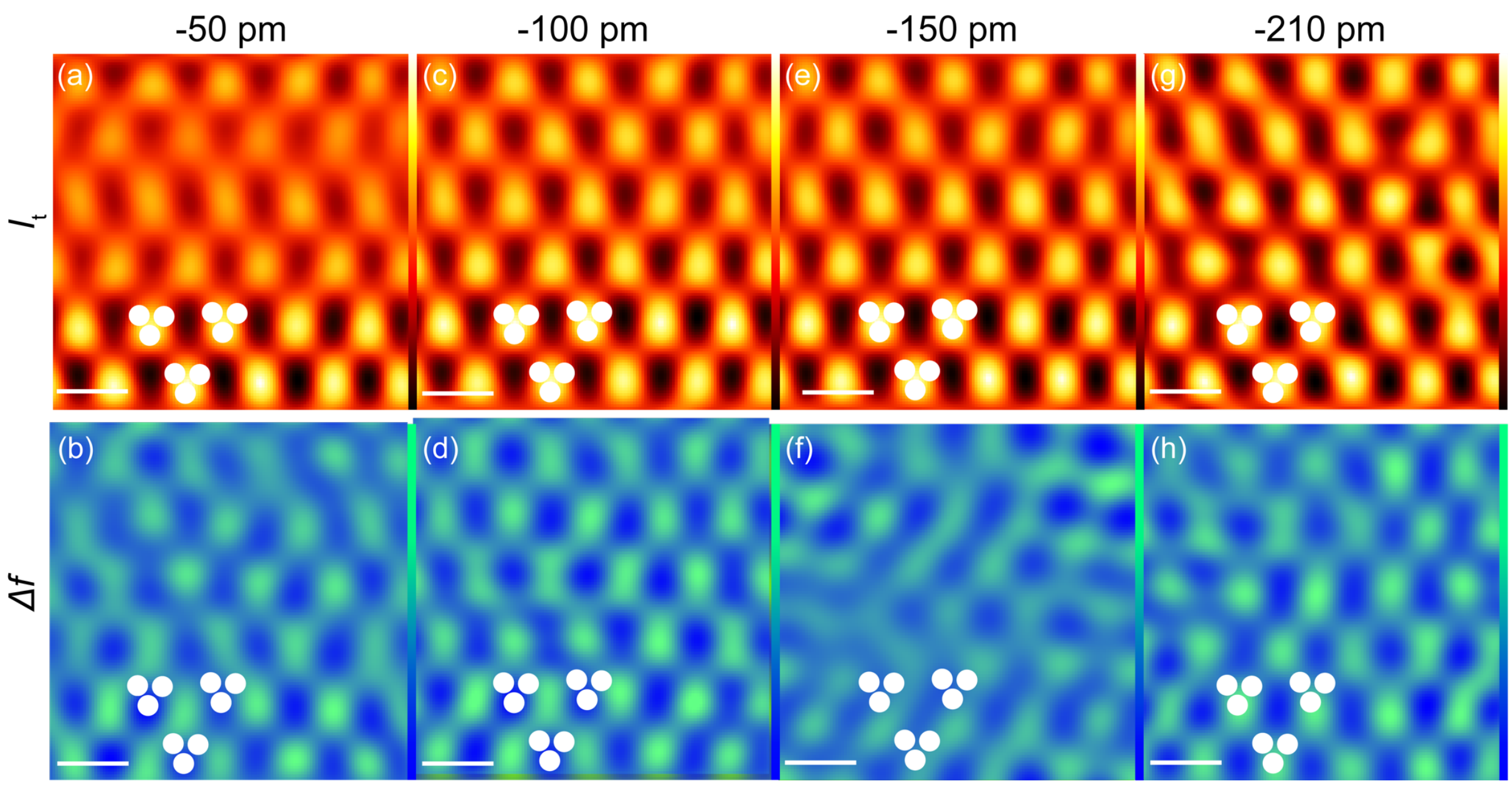

Figure S5: Inverse FFT of constant-height $I_t$ and $\Delta f$ images at indicated tip displacements $z$ ($V_s$ = 3 mV, $A_{osc}$ = 50 pm). $I_t$ and $\Delta f$ images have been row corrected before performing the FFT. The tip has been stabilized at $V_s$ = 100 mV, $I_t$ = 10 pA. Scale bars are 1 nm in all images.

To illustrate the relative position of the CDW maxima in constant-height $I_t$ and $\Delta f$ images at different $z$, Fig. S5 presents filtered inverse FFT images of the CDW of the data shown in Fig. S2. At small tip-sample distances ($z$ = -50 pm), the maxima in the $I_t$ image correspond to the minima in the $\Delta f$ image, which means that the regions of largest current correspond to the strongest attractive force. At more negative $z$, this relation changes. At $z$ = -210 pm, the maxima in the $I_t$ image do not correspond to the positions of largest attraction in the $\Delta f$ image anymore.

## Section S6: Analysis procedure of FFT images

To obtain the FFT peak intensity ($A^{\mathrm{FFT}}$) of the atomic lattice ($q_{\mathrm{atom}}$) and CDW ($q_{\mathrm{CDW}}$) from row-corrected constant-height $I_{\mathrm{t}}$ and $\Delta f$ images at different $z$, FFT images are obtained using the FFT transformation in Matlab. The full analysis procedure is illustrated in Figure S6.

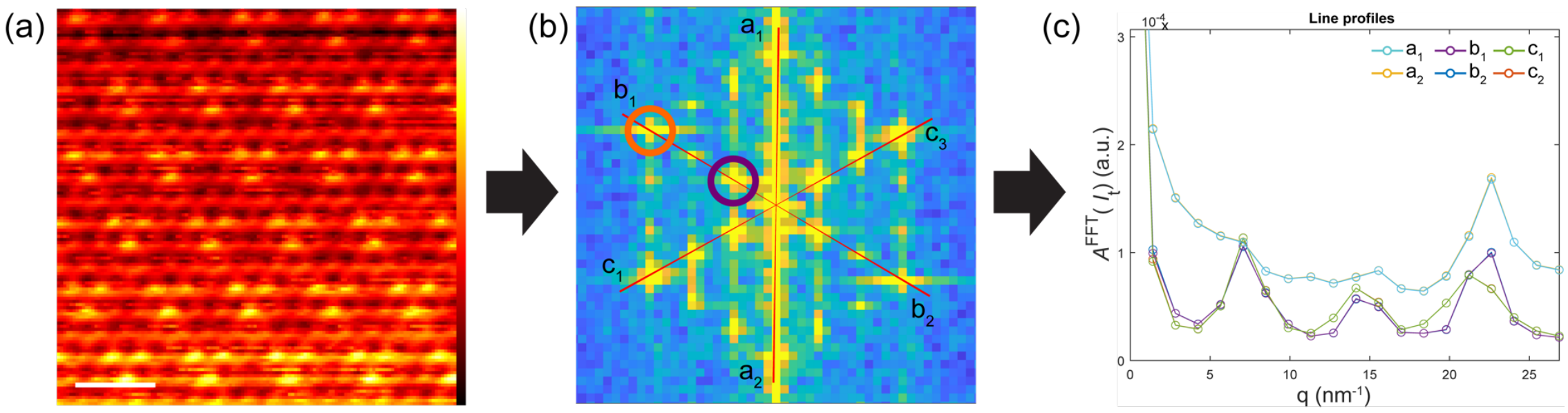

Figure S6: Procedure for obtaining the FFT intensities $A^{\mathrm{FFT}}$. For a selected image (a), the FFT transformation is obtained using Matlab (b). Line profiles (c) averaged over 20 separate lines as indicated in the image are obtained for the high-symmetry axes and the peak values at $q_{\mathrm{atom}}$ and $q_{\mathrm{CDW}}$ are extracted (marked by orange and violet circles in (b)).

In the FFT image (Fig. S6(b)), the center peak is suppressed to a threshold value in order to maximize the contrast for $q_{\mathrm{atom}}$ and $q_{\mathrm{CDW}}$. This value was kept the same throughout the entire data set. Line profiles are extracted from the FFT image by an average over the span given by the red lines in Fig. S6(b). The averaged line profiles along the high-symmetry directions are shown in Fig. S6(c). From the average value of these profiles, are extracted for $q_{\mathrm{atom}}$ and $q_{\mathrm{CDW}}$. The error margins are defined by the standard deviation. Depending on the orientation of the images, the fast scan direction can align with the FFT peaks on the high-symmetry axis. In such a case only two axes can be analyzed. We note that this had no influence on our findings discussed in the following.

Figure S7 shows $A^{\mathrm{FFT}}$ of $q_{\mathrm{atom}}$ and $q_{\mathrm{CDW}}$ extracted from constant-height $I_t$ and $\Delta f$ images versus $z$ for various data sets that have been acquired with 4 different tips, different orientations of the image frame to the high-symmetry axes and different oscillation amplitudes $A_{\mathrm{osc}}$ of 50 pm and 60 pm. For $I_t$ images, $A^{\text{FFT}}(z)$ follows a rather exponential increase for more negative tip displacements for both $q_{\text{atom}}$ and $q_{\text{CDW}}$ as indicated by the fitted dashed lines. One exception is the tip 3 (Fig. S7(e)) which deviates from the trend at the most negative tip displacement. The slope and absolute values differ depending on the tip. Still, both $q_{\text{atom}}$ and $q_{\text{CDW}}$ exhibit similar intensities for the entire probed $z$ range for all tips. We note that for most tips, $I_t(z)$ deviates from the expected exponential behavior at more negative $z$ as indicated by the fitted dashed lines in Figs. S7(i-l). We assign this to relaxations of the tip or the surface at smaller tip-surface distances. The relaxations do not appear to have a significant influence on the observed trends of $A^{\mathrm{FFT}}(z)$ for $q_{\mathrm{atom}}$ and $q_{\mathrm{CDW}}$ though it may be the cause for the deviation from exponential behavior for tip 3. $A^{\mathrm{FFT}}(z)$ extracted from constant-height $\Delta f$ images show an exponential increase of $q_{\mathrm{atom}}$ with more negative $z$ while no clear trend for $A^{\mathrm{FFT}}(z)$ for $q_{\mathrm{CDW}}$ is observed.

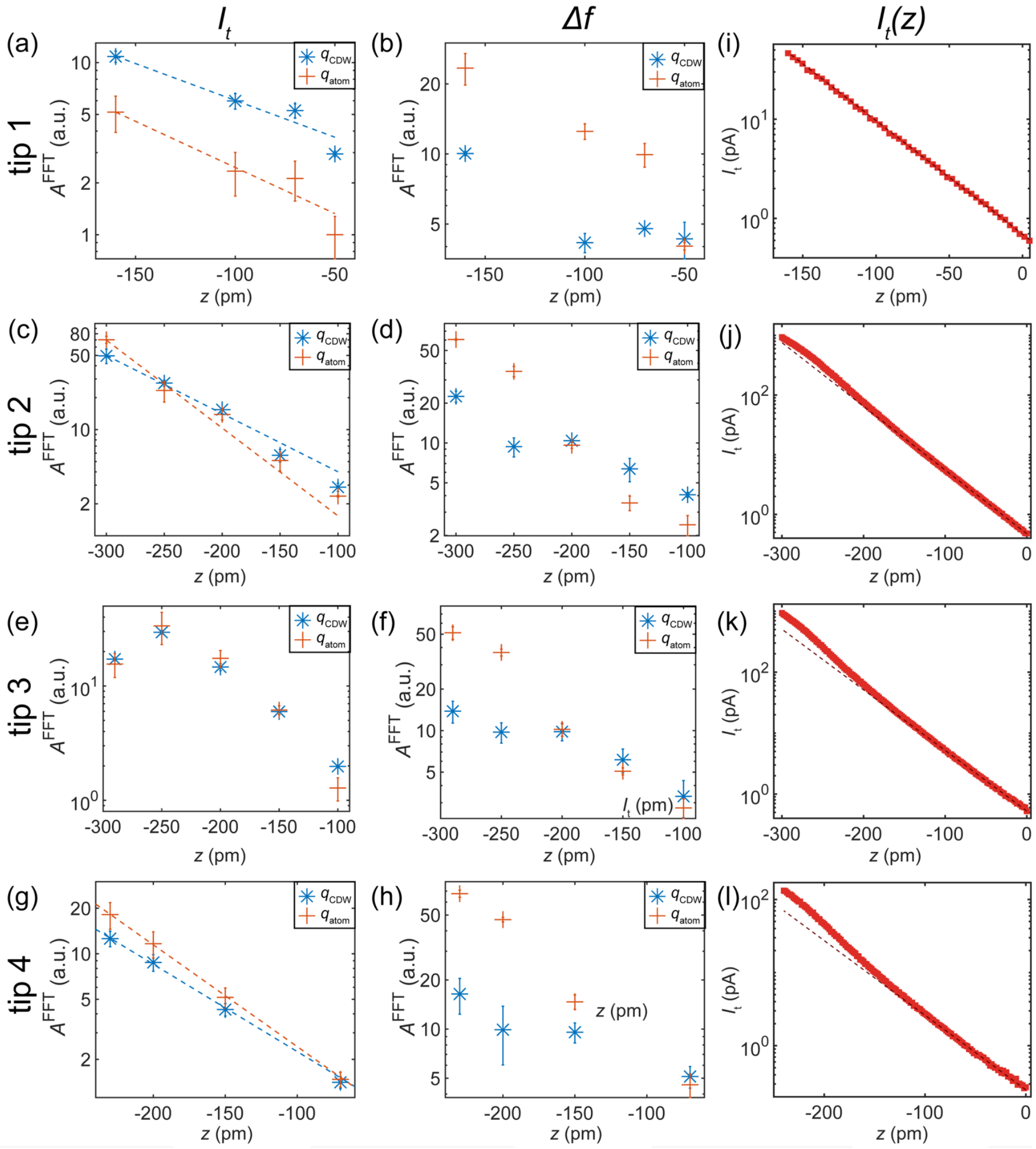

Figure S7: (a-h): Logscale FFT peak intensity ($A^{FFT}$) of the atomic lattice ($q_{\text{atom}}$) and the CDW ($q_{\text{CDW}}$) extracted from constant-height images of the tunneling current $I_t$ (left column) and of the frequency shift $\Delta f$ (middle column) versus tip displacement $z$ for four different tips. (i-l) $I_t$ versus $z$ acquired with the respective tips. The marker size indicates the error bar. ($V_s$ = 100 mV, $I_t$ = 10 pA; tip 1: $A_{\text{osc}}$ = 50 pm; tip 2: $A_{\text{osc}}$ = 60 pm; tip 3: $A_{\text{osc}}$ = 60 pm; tip 4: $A_{\text{osc}}$ = 60 pm).

## Section S7: STS measurements

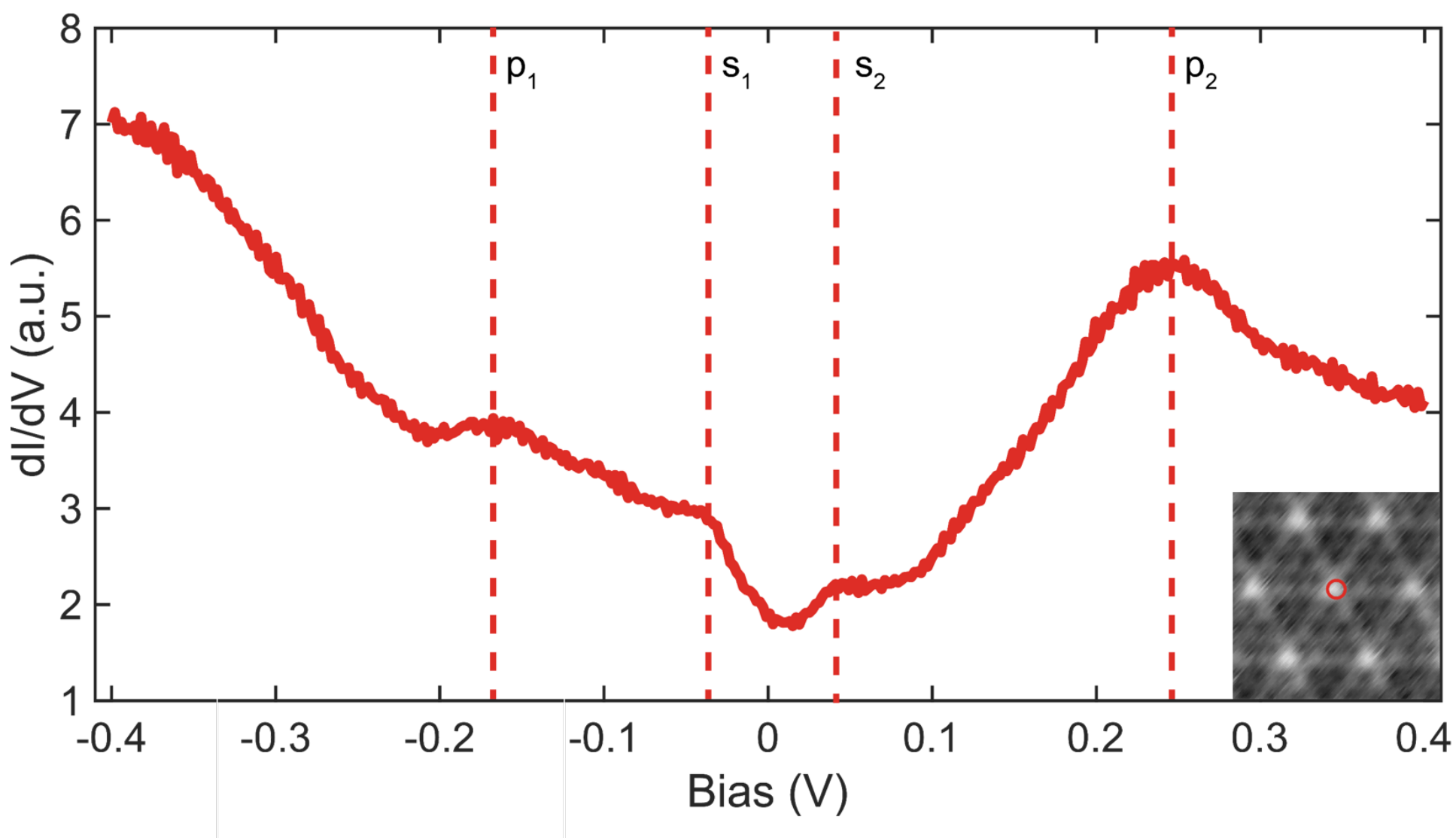

Figure S8: Representative constant-height $dI_t/dV$ spectrum versus $V_s$ obtained at the position as marked in inset (tip stabilization: $V_s = 40$ mV, $I_t = 200$ pA, $V_{\mathrm{mod}} = 5$ $\mathrm{mV_{rms}}$).

Figure S8 shows a $dI_t/dV(V_S)$ spectrum at a fixed tip-surface distance above the CC region (position indicated by the empty circle in the inset STM topography). We identify two shoulders $s_1$ and $s_2$ at -47 ± 15 mV and 42 ± 7 mV, respectively, in agreement with previous studies [9,10]. The error margin is given by the spread of the peak positions as measured with different tips. In the past, $s_1$ and $s_2$ were interpreted as the CDW energy gap [9,11]. However, they are also present above the critical temperature for the CDW phase and have recently been ascribed to result from inelastic tunneling processes [12-14]. To date, no convincing signature of the CDW gap in STS measurement for bulk 2H-$NbSe_2$ was reported. This is ascribed to the fact that the CDW gap originates from bands at the $M$ point within the Brillouin zone whereas STS is mainly sensitive to tunneling processes at the Γ point [12]. We note that for a single layer of 2H-$NbSe_2$ a clear energy gap was observed [15]. Our measured $dI_t/dV(V_S)$ spectrum also exhibits two peaks ($p_1$ and $p_2$) at -149 ± 34 mV and 228 ± 46 mV, similar to what has been reported before in both the CDW and normal phase [14]. Overall, our STS spectrum agrees well with those reported previous studies.

## Section S8: Layer-dependent electronic structure

Figure S9 displays the crystal structures which compare well with the previously reported structures [16-19]. The reference point for the distance *d* above the surface to the topmost Se plane is defined as the mean out-of-plane position of the Se ions within that plane (dashed line in S9(c)).

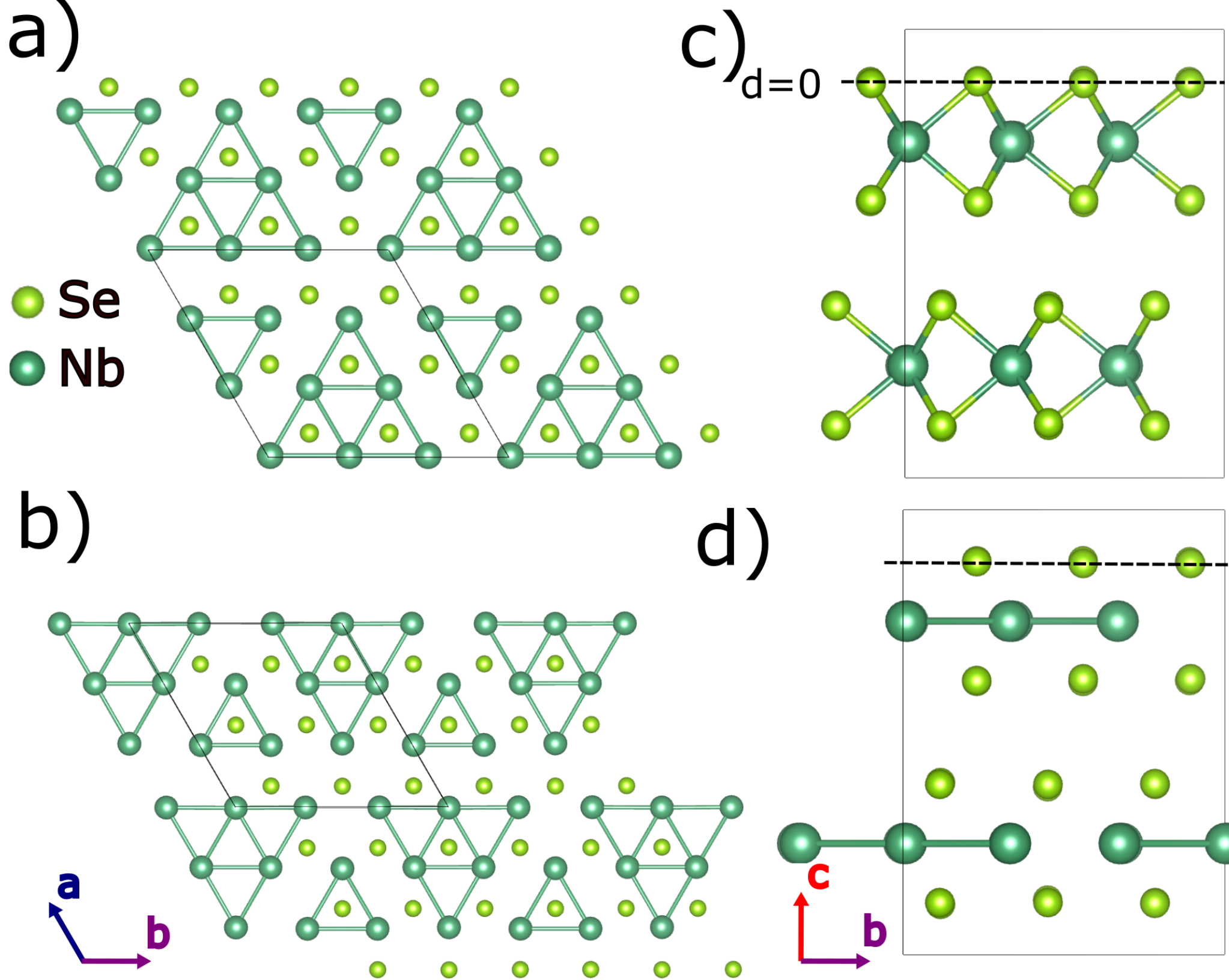

Figure S9: Relaxed 3 x 3 supercell crystal structures. (a) HC top view, (b) CC top view. Green lines indicate a shorter distance between the Nb atoms than the equilibrium distance. (c), (d) different side views of CC-HC-CC-HC stacking between layers. Green bonds between Niobium atoms illustrate which Nb atoms move close together in the CDW. The position of $d = 0$ pm is indicated in (c). Structures visualized using VESTA [5].

Figure S10 displays the obtained band structures and DOS of the structures with 1 to 4 layers. Within the increasing number of layers, the electronic structure is expected to more closely resemble that of bulk 2H-NbSe$_2$ [16,20]. In particular, we observe that a $p_z$ dominated band moves up in energy around Γ as we increase the number of layers, in agreement with a previous work [21]. This band is strongly $k_z$-dispersive in bulk as shown in Fig. S10(e). Consequently, the energy position of the $k_z$-dispersive band can strongly change depending on the number of layers in the simulated DOS and d$I_{sim}$/d$V$ images, especially for energies further away from the Fermi level. In particular, the spatial positions of maximum intensity in d$I_{sim}$/d$V$ images are different between the trilayer (Figs. S11(a) and (c)) and 4-layer model (Figs. S11(b) and (d)) at

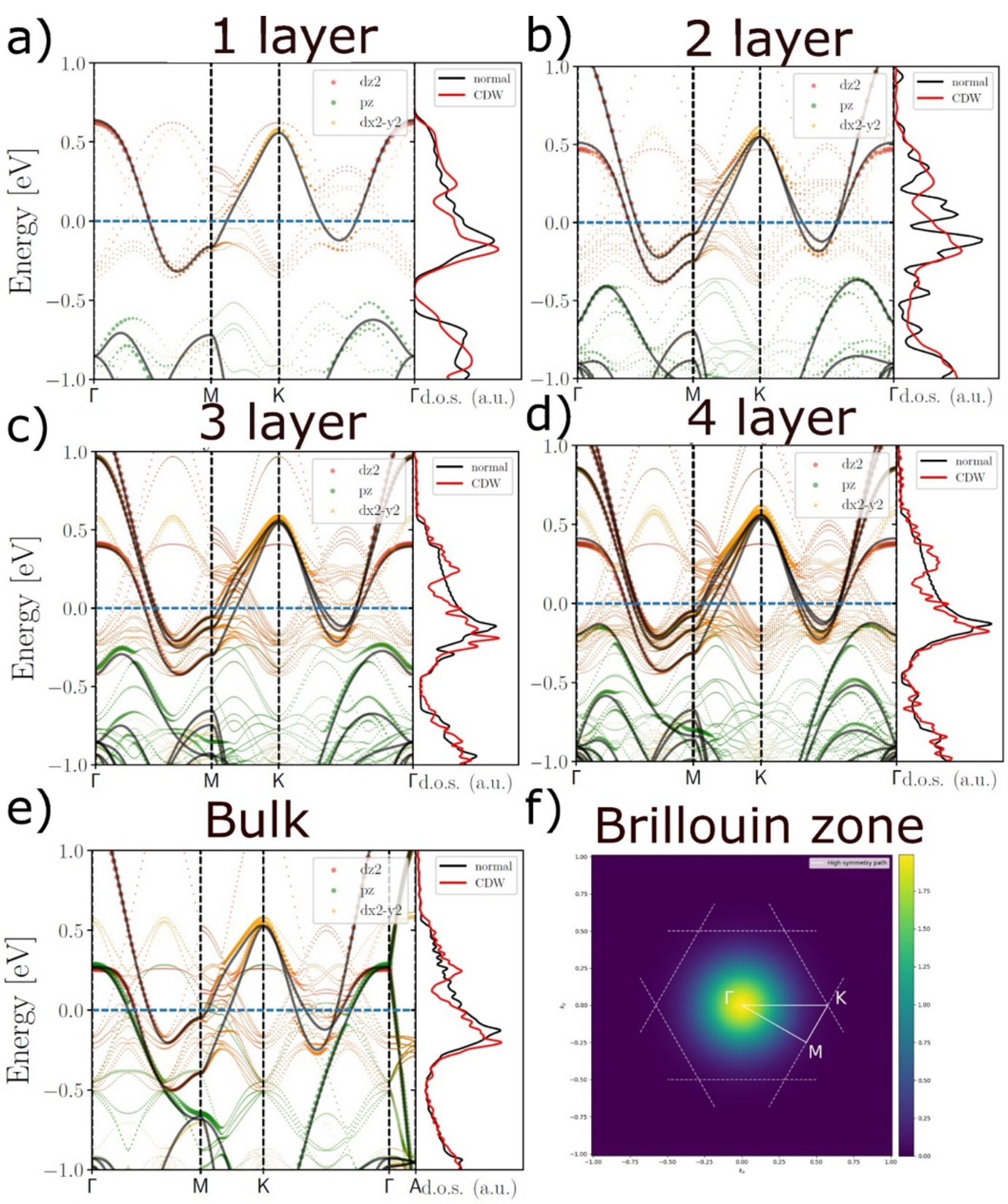

Figure S10: Unfolded band structures of 2H-NbSe$_2$ structures for 1 to 4 layers and bulk in the CDW phases. The bulk band structure in (e) depicts the cut along Γ-M-K-Γ at $k_z = 0$. Top view of 1st Brillouin zone with high symmetry path marked. The tip-sample tunneling matrix $T(k_x, k_y)$ used to same energies. The contrast in the $dI_{sim}/dV$ image for the trilayer model at -250 meV (Fig. S11(c)) is similar to that of the 4-layer system for -200 meV (Fig. S11(b)). We assign this to the strong influence of the $p_z$ band that enters the respective energy window.

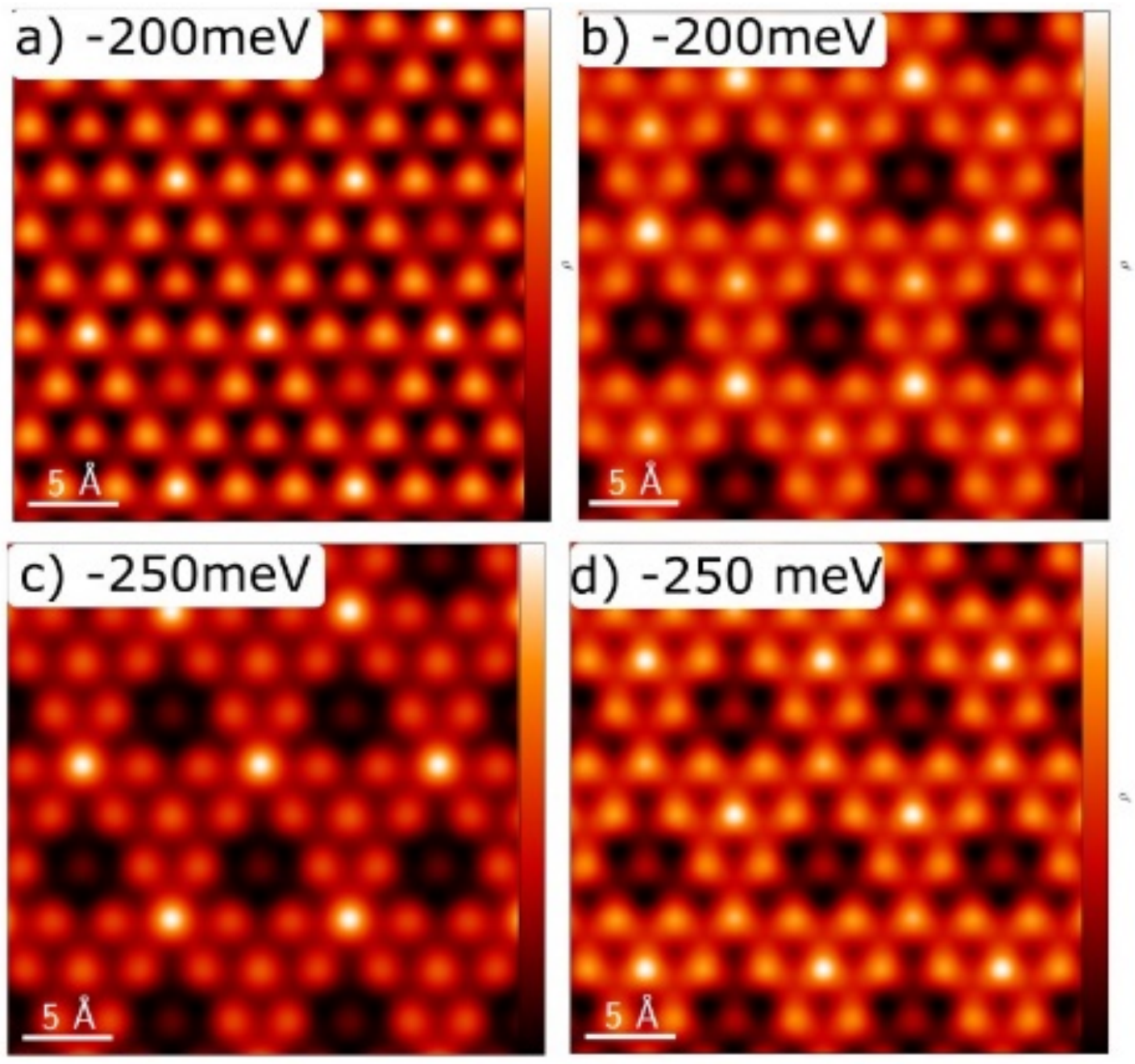

Figure S11: Simulated $dI_{sim}/dV$ at elevated voltages for the (a), (c) trilayer and (b), (d) 4-layer models at $d$ = 320 pm from the Se plane and indicated voltages for the CC structure.

## Section S9: Calculated density of states for bulk 2H-$NbSe_2$

We model the sensitivity of the tunneling current to energy states around the Γ-point by letting the tunneling matrix be $T(k_x, k_y) = \left(1/(\sigma_T\sqrt{2\pi})\right) exp\left(\frac{-(k_x^2+k_y^2)}{(2\sigma_T^2)}\right)$. $T(k_x, k_y)$ is plotted in the Brillouin zone (Fig. S10(f)). The calculated DOS shown in Fig. S12 is projected onto the atomic states of the same Se ion for which the STS spectrum in Fig. S8 was measured, and is shown for four different values of $\sigma_T$. Given that a Gaussian tunneling matrix T with $\sigma_T$ = 0.2 |b| gives the closest qualitative match between measured (Figs. S8) and calculated STS for bulk $NbSe_2$, we choose this value for our calculations in the trilayer model. In Fig. S12, we identify a dominant peak at -0.18 eV. This peak compares well with the position of $p_1$ observed in the measured $dI_t/dV(V_s)$ (Fig. S8). We further find a peak at about +0.3 eV which may be comared to $p_2$ in Fig. S8. We interpret this peak to mainly originate from the strongly $k_z$-dispersive $p_z$ band. The features around the Fermi level here do not compare well with the shoulder $s_1$ and $s_2$ in Fig. S8. However, as $s_1$ and $s_2$ have been attributed to inelastic tunneling processes [12-14], we do not expect to observe these in the calculated electronic DOS.

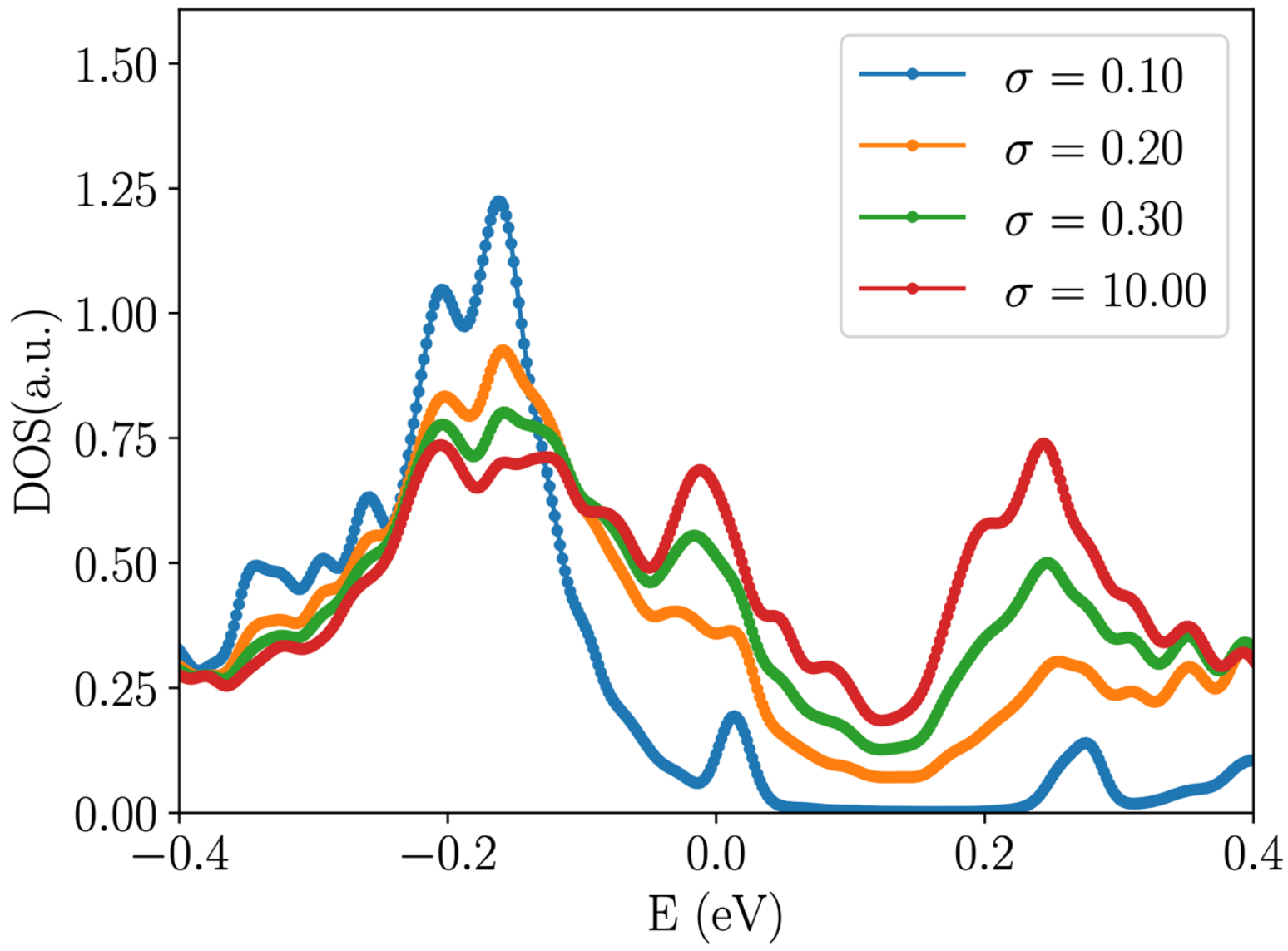

Figure S12: Density of states (DOS) versus energy $E$ for bulk 2H-$NbSe_2$. The DOS is projected onto the atomic states of the same Se ion for which the STS spectrum in Fig. S8 was measured. In addition, the DOS is weighted around the Γ point by the tip-sample tunneling matrix $T$ for different values of $\sigma_T$. The spatial extension of $\sigma_T$ = 0.2 |b| is plotted in Fig. S10(f). $\sigma_T$ = 10 |b| corresponds to the projected density of states with a constant $T$.

## Section S10: Images of the differential conductance for CC and HC regions

Figures S13(a-c) show measured $dI_t/dV$ images where both CC and HC dominated regions are present as marked by the dashed and solid circle, respectively. The images at -50 mV and +40mV show a reasonable agreement for both the HC and CC structure. However, there is a deviation regarding the position of maximum contrast between the experimental and simulated images at -250 mV. We ascribe this to the fact that our trilayer model does not correctly describe the energetic position of the $p_z$ band close to the Γ point, in comparison to bulk 2H-$NbSe_2$. Altogether, we conclude that our calculated Γ-weighted DOS for the trilayer model shows a decent agreement with the experimental findings for energies within -50 meV and +40 meV, and we use this model for the further analysis of the distance-dependent evolution of $q_{atom}$ and $q_{CDW}$.

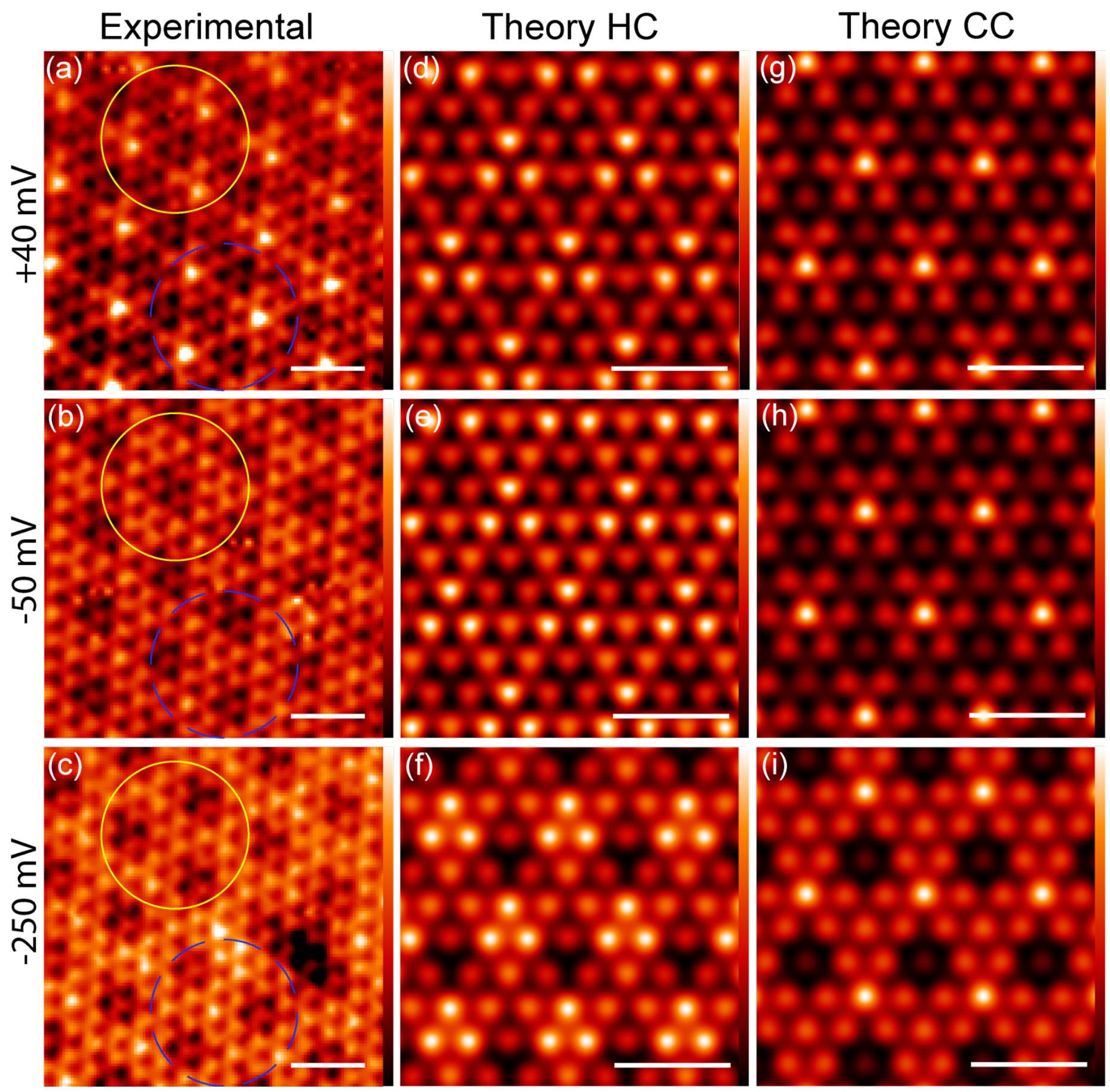

Figure S13: Measured $dI_t/dV$ images recorded at (a) $V$ = +40 mV, (b) -50mV and (c) $V$ = -250 mV. The solid and dashed circles mark the regions that are close to HC and CC, respectively. Tip stabilization: $V_s$ = 50 mV, $I_t$ =100 pA, $V_{mod}$ = 5mV$_{rms}$ above an HC region. (a), (b) and (c) are row corrected and gaussian filtered with 5 points with the color scale adjusted for maximum contrast. Scale bars = 1 nm. Simulated $dI/dV$ images at (d) $V$ = +40 mV, (e) -50mV and (f) $V$ = -250 mV at $d$ = 320 pm above a HC region and (g), (h), (i) above a CC region.

## Section S11: Distance-dependent calculations for CC regions

The calculated distance-dependent behavior of $q_{CDW}$ and $q_{atom}$ show the same trends for the CC structure as for the HC structure as depicted in Figs. S14 and S15.

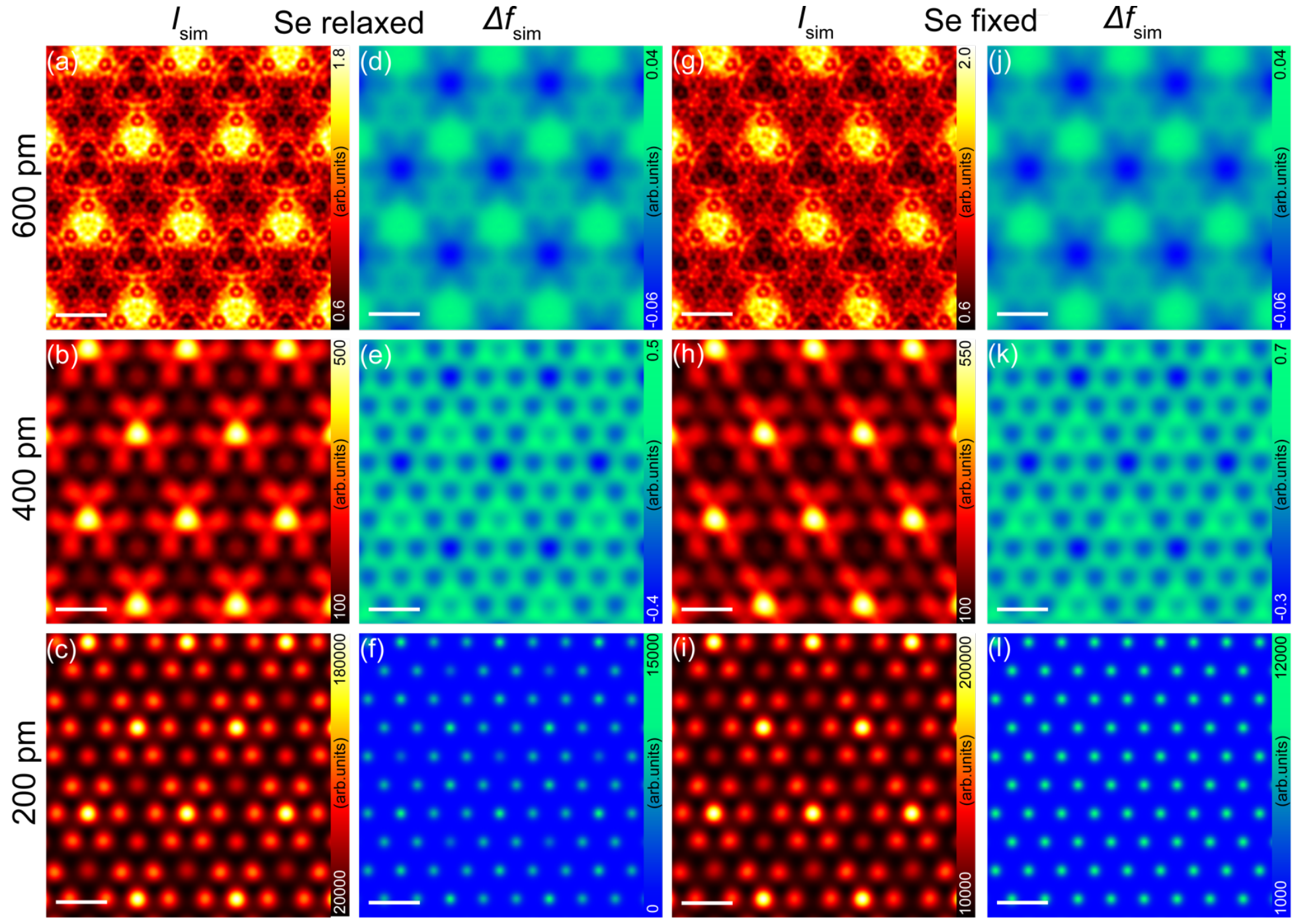

Figure S14: (a-c) Simulated current $I_{sim}$ and (d-f) simulated frequency shift $\Delta f_{sim}$ images for a CC region for different distances $d$ for the relaxed structure. (g-l) Same as in (a-f) but for the fixed structure. The current images were calculated using states between energy states between $E_F$ and $E_F$ + 50 meV. $A_{osc}$ = 50 pm. Scale bar is 0.5 nm in all images.

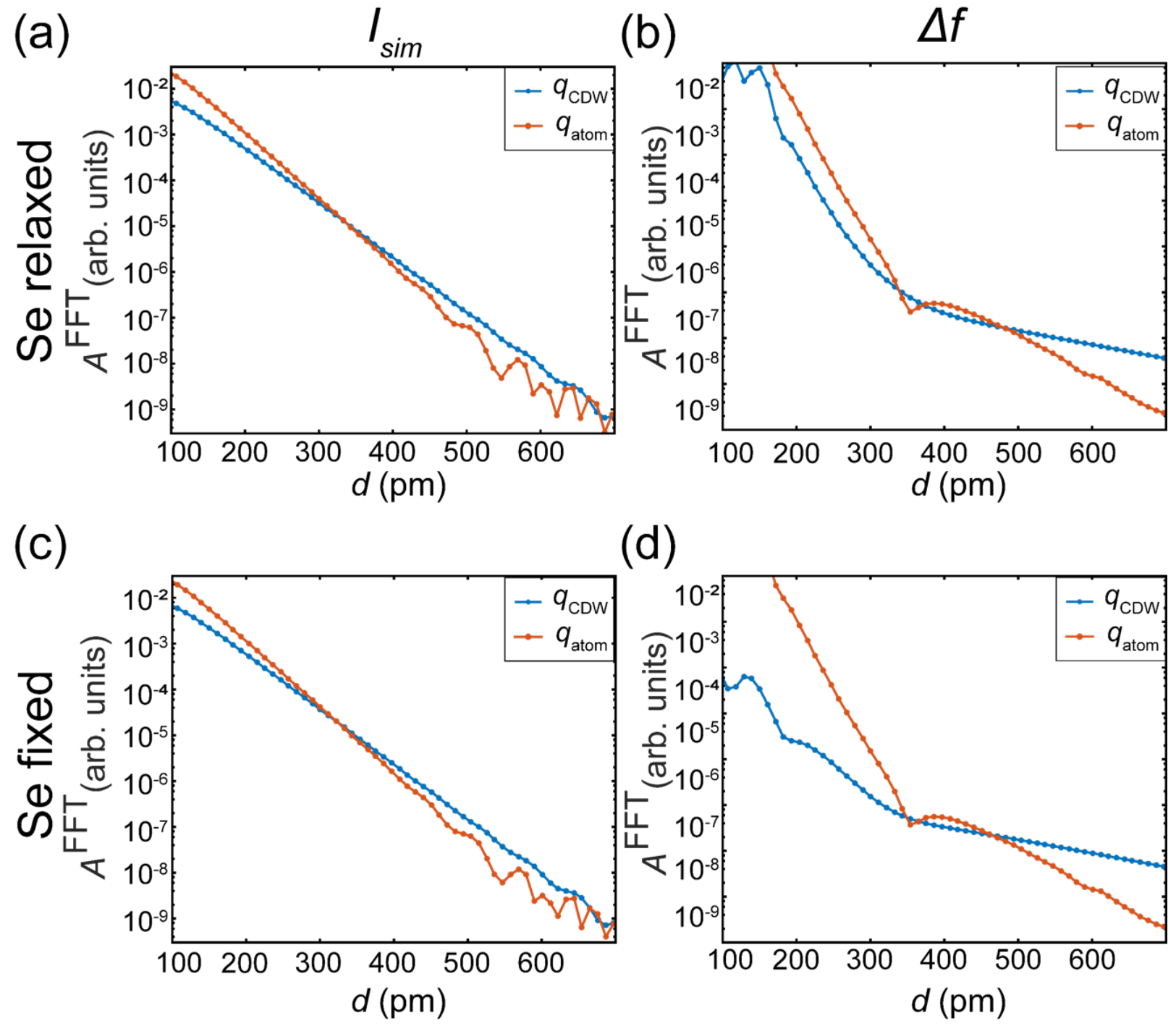

Figure S15: (a, b) Semi-log plot of FFT peak intensity ($A^{FFT}$) of $q_{atom}$ and $q_{CDW}$ vs. $d$ extracted from $I_{sim}$ (a) and $\Delta f_{sim}$ (db images for the relaxed structure. (c, d) Same as (a, b) but for the structure with fixed Se atoms. All images were measured or calculated for a CC region.